\documentclass[]{jfm_arxiv}
\usepackage{graphicx}
\usepackage{newtxtext}
\usepackage{newtxmath}
\usepackage{natbib}
\usepackage{subfig}
\usepackage{setspace}
\usepackage{hyperref}
\hypersetup{
    colorlinks = true,
    urlcolor   = blue,
    citecolor  = blue,
}

\newcommand{\RomanNumeralCaps}[1]
\linenumbers
\title{Compressibility Driven Wake Transition and Hysteresis over Cargo Aircraft Aftbodies}
\author{Chitrarth Prasad\aff{1}
  \corresp{\email{c.prasad@okstate.edu}}, Rajesh Ranjan\aff{2}, Daniel J. Garmann\aff{3}       \and Datta V. Gaitonde\aff{4}}
\affiliation{\aff{1} School of Mechanical and Aerospace Engineering, Oklahoma State University, OK 74078
\aff{2} Department of Aerospace Engineering, Indian Institute of Technology Kanpur, India
\aff{3} Air Force Research Lab, Wright Patterson AFB, OH 45433
\aff{4} Department of Mechanical and Aerospace Engineering, The Ohio State University, OH 43210}

\usepackage{xargs}
\usepackage[colorinlistoftodos,prependcaption,textsize=tiny]{todonotes}
\newcommandx{\unsure}[2][1=]{\todo[linecolor=red,backgroundcolor=red!25,bordercolor=red,#1]{#2}}
\newcommandx{\needswork}[2][1=]{\todo[linecolor=blue,backgroundcolor=blue!25,bordercolor=blue,#1]{#2}}
\newcommandx{\info}[2][1=]{\todo[linecolor=OliveGreen,backgroundcolor=OliveGreen!25,bordercolor=OliveGreen,#1]{#2}}
\newcommandx{\improvement}[2][1=]{\todo[linecolor=Plum,backgroundcolor=Plum!25,bordercolor=Plum,#1]{#2}}
\newcommandx{\thiswillnotshow}[2][1=]{\todo[disable,#1]{#2}}
\usepackage{ulem}
\usepackage{soul}

\begin{document}
\maketitle

\begin{abstract} 
Aft sections of military cargo aircraft employ flat surfaces at high upsweep angles to accommodate ramp doors,  producing flow features that affect cargo-drop accuracy, paratrooper safety, and aerodynamic performance.
Fundamental studies have primarily examined near incompressible flow over a canonical surrogate consisting of a freestream aligned cylinder with a planar, sharp edged upswept base.
The flow exhibits peripheral separation, a horseshoe vortex, and a counter-rotating streamwise vortex pair that persists downstream.
%Key flow features include peripheral separation, a horseshoe-type vortex, and a counter-rotating streamwise vortex pair that persists downstream.
The present investigation delineates the effects of compressibility on the wake and examines how these effects depend on basal upsweep angle.
Wall-resolved large-eddy simulations are performed at Mach numbers of $0.1$, $0.3$, and $0.5$ for upsweep angles of $32^\circ$ and $45^\circ$ at a nominal Reynolds number of $25{,}000$.
For the $32^\circ$ afterbody, increasing Mach number enlarges the upstream recirculation region and delays vortex-pair formation, while these effects diminish downstream.
For the $45^\circ$ afterbody, similar recirculation-region growth triggers a bifurcation at Mach~0.5 from the vortex-pair state to a broad separated turbulent wake. 
A descending-Mach sequence to 0.3 and 0.1 reveals hysteresis, with the separated-wake state persisting at lower Mach numbers and remaining robust to Reynolds-number variation. 
%Thus, both states can occur at identical Mach and Reynolds numbers. 
%Compressibility therefore acts as a wake-state-selection mechanism, making flow topology and surface-pressure loading dependent on Mach-number history.
Thus, both states can occur at identical Mach and Reynolds numbers, with topology and pressure loading governed by Mach number history.

%Upsweep angle, flow characterized by streamwise vortex pair.
%Surrogate configuration.
%Only consider incompressible flow.
%Emphasis on mean flow.

%\noindent\textbf{Keywords:} cargo aircraft, vortex-pair, drag, hysteresis
\end{abstract}

\begin{keywords}
%turbulence modeling, hypersonic
\end{keywords}

%\end{frontmatter}

%\linenumbers
\section{\label{sec:Intro} Introduction}

%Why are cargo aircraft important to the airforce.
%Military transport aircraft serve a crucial function in a wide range of strategic and tactical missions %by providing rapid transportation of personnel, supplies, and equipment
%including aerial refueling, airlifting, and rescue operations in remote and difficult-to-reach locations.
Military transport aircraft play an essential role in supporting military operations by providing rapid transportation of personnel, supplies, and equipment to areas in need.
%To support these missions, 
Due to their specialized function, military transport aircraft have unique rear fuselage designs that distinguish them from conventional aircraft. 
The two key design differences include: (i)~a~large upsweep angle and (ii) a relatively flat base; these features collectively enable the installation of ramp doors for the loading and unloading of cargo and the deployment of paratroopers. 

%Results in vortex pair
%A direct consequence of these unique design features is the formation of a %result in flow separation around the periphery of the flat base, the formation of a 
%Although the abovementioned design features are essential for the effectiveness of military transport aircraft, they also create some aerodynamic challenges. %in terms of aerodynamics and stability.
Model-scale experiments~\citep{epstein1994experimental,bury2013experimental} have shown that these distinctive design features give rise to a streamwise-oriented vortex pair that dominates the afterbody dynamics. %
This vortex pair generates strong inboard and vertical induced velocities, increasing the risk of parachute collisions, tail strikes, and inaccurate cargo drops.  
%This highly unsteady longitudinal vortex pair
%can interfere with cargo drops and paratrooper activity, and raise the possibility of tail strikes.
In addition, the low-pressure footprint of this vortex pair on the flat base surface reduces fuel efficiency by contributing to the total drag of the aircraft~\citep{wortman1999reduction,smith2013reduction,telli2016investigation,carter2017legacy}. 
For instance, studies estimate that the vortex pair can account for as much as $11\%$ of the total drag experienced by the Lockheed C-130~\citep{smith2013reduction}. 
Furthermore, the strong streamwise vortex pair core exhibits large fluctuations downstream, known as meandering~\citep{wang2012unsteady,zigunov2020dynamics,ranjan2020meandering}, potentially increasing the safe longitudinal separation distance from other aircraft, especially during take-off and landing stages.
A comprehensive understanding of the flow is, therefore, crucial to improving cargo aircraft aft-body design.

\begin{figure}
    \centering
%    \subfloat[]{\includegraphics[width=0.5 \textwidth]{Compressibility_Effects/Figures/FullGeom_3DView.jpeg}} \\
%        \subfloat[]{\includegraphics[width=0.75 \textwidth]{Compressibility_Effects/Figures/FullGeom_ZView.jpeg}}
\includegraphics[width=\textwidth]{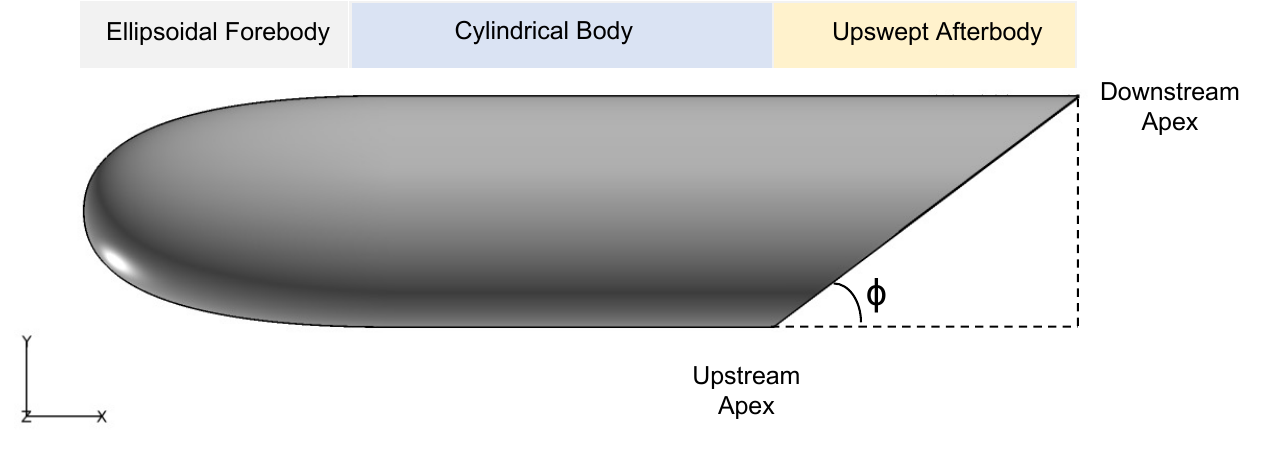}
    \caption{Canonical slanted-base cylindrical afterbody used in this work.}
    \label{fig:FullGeom}
\end{figure}

Figure~\ref{fig:FullGeom} presents a canonical surrogate configuration often used to understand the dynamics behind cargo aircraft. %study such flows.
This configuration consists of a cylinder with an ellipsoidal forebody placed parallel to the freestream flow and truncated at an angle (the upsweep angle, $\phi$) to form a proxy aft-body.
The points designated upstream and downstream apexes are marked for subsequent reference. 
%Bulatsinghala et al.~
\cite{bulathsinghala2017afterbody} conducted flow measurements for this surrogate geometry for a range of upsweep angles and found a $50\%$ increase in drag coefficient from $\phi = 24^\circ$ to $\phi = 32^\circ$.
This drag increase was accompanied by an increase in vortex circulation and a subsequent decrease in the meandering amplitude of the vortices near the downstream apex.
%Garmann and Visbal~
\cite{garmann2019high} performed highly-resolved large eddy simulations (LES) at $\phi=28^\circ$ using higher-order methods to capture the fine-scale features of the vortex pair measured by~\cite{bulathsinghala2017afterbody}. 
Their simulations highlighted regions of secondary vorticity, similar to ones encountered in delta wings~\citep{gursul2005unsteady,gordnier2009computational}, near the base surface that were not captured previously in the experiments.
These secondary structures were successfully captured by~\cite{zigunov2020reynolds} using stacked stereoscopic particle image velocimetry (S-SPIV) measurements and oil flow visualizations at different Reynolds numbers ($Re$) and $\phi$ values.
In addition, their observations indicate the existence of a bi-stable state defined by either a `vortex-pair' or a `separated-wake' closure at $\phi=45^\circ$, consistent with the earlier experiments of~\cite{morel1978effect,morel1980effect} and~\cite{britcher1991interference}.
This change in flow regime at high upsweep angles was numerically captured by %Ranjan et al.
~\cite{ranjan2020mean} by performing a series of LES simulations at $\phi$ values ranging from $20^\circ$ to $55^\circ$.
Their results suggest a potential pathway by which the streamwise-aligned vortex-pair may ultimately
result into a turbulent wake-type regime.
These insights were subsequently leveraged by %Ranjan and Gaitonde
~\cite{ranjan2020hysteresis} to numerically explore the role of hysteresis in the manifestation of the two regimes.

While these investigations have yielded several important insights into the flow downstream of transport aircraft fuselages, most focused on relatively incompressible conditions, whereas the cruise speeds of typical cargo aircraft, range from Mach~$0.5$ to $0.6$.  
%As such, compressibility effects on the flow can be significant.
A notable exception to the focus on incompressible speeds, is a recent effort by~\cite{huss2025compressibility}, where the Mach number was varied from $0.3$ to $0.6$ for both sharp-edged as well as rounded bodies.  
%The increased  recirculation region with Mach number shown to be consistent with vortex compression effects.
%Even though the near base flow was affected, the vortex circulation for the rounded body was nearly invariant with Mach number. 
%Under these conditions, various density-related effects can arise, especially in the relatively lower pressure wake region.
Increasing Mach number was found to reduce shear-layer growth and entrainment, producing a larger recirculation region.
For the rounded-edge configuration, substantial changes in the near-base flow occurred while the downstream vortex circulation remained nearly invariant with Mach number.
The sharp-edge configuration exhibited a more pronounced response, transitioning from a vortex-dominated state at Mach~0.3 to a fully separated wake at Mach~0.6.

%Under these conditions, various density-related effects can arise, especially in the relatively lower pressure wake region.
These observations are consistent with the broader influence of compressibility on separated and vortical flows.
In general, compressibility can modify separation
patterns~\citep{presz1974flow,presz1976analytical}, the growth and
stability of shear layers~\citep{sandham1990compressible,
sandham1994effect,leep1993three}, and the evolution of vortical
structures~\citep{bohorquez2013three,ohmichi2017compressibility}.
Such effects have been documented over delta
wings~\citep{riou2010compressibility,luckring2004compressibility} and
axisymmetric forebodies~\citep{keener1975wind}, both of which, like the
afterbody configuration considered here, exhibit prominent streamwise
vortices and separation-induced wakes.
%\textbf{Add some stuff about compressibility on delta wings, tip vortices, etc.}
%Self and mutually induced effects on the vortices are also relevant. 
More practical manifestations include changes in drag and lift coefficients~\citep{melin2010induced,von2003compressibility,wootton1967effect}.
%To highlight
%the changes in the vortical wake due to potential compressibility effects

This investigation builds on these observations by leveraging wall-resolved LES to delineate the effect of compressibility, including the influence of initial conditions, on the %genesis and evolution of the vortex-pair 
wake of slanted base cylindrical afterbodies shown in Fig.~\ref{fig:FullGeom}.
To achieve this, we perform multiple sequences of wall-resolved LES simulations at increasing Mach numbers in the range $0.1$ to $0.5$ at upsweep angles of $32^\circ$ and $45^\circ$.
%The goal of this investigation is to delineate the effect of compressibility on the %genesis and evolution of the vortex-pair 
%wake of slanted base cylindrical afterbodies shown in Fig.~\ref{fig:FullGeom}.
%This is achieved by performing multiple sequences of well-resolved LES simulations with increasing Mach numbers in the range 0.1 to 0.5 at upsweep angles of $32^\circ$ and $45^\circ$.
The LES simulations build on the previous results at Mach~$0.1$ of %Ranjan et al.
~\cite{ranjan2020mean} and %Ranjan and Gaitonde
~\cite{ranjan2020hysteresis} which were performed to collectively examine the effect of upsweep angle and \textit{Re} on the flow evolution.
%\textbf{The present investigation uses two upsweep angles ($\phi_1=32^\circ$ and $\phi_2=45^\circ$); Need to write nicely and connect this to vortex and wake regimes} %and is kept constant at the 
%$\phi_1$ is very close to the upsweep angle of C-130 Hercules ($\phi \approx 28^\circ$) whereas $\phi_2$ serves as .
%A summary of the LES methodology is presented in $\S$~\ref{sec:LES}, including the
%The LES 
%simulation sequences used to establish the effect of compressibility. %are introduced in $\S$~\ref{sec:LES2}.
%The change in mean flow topology, vortex evolution and base surface pressure for the 32^
The numerical methodology and the simulation sequences used to isolate
the influence of Mach number are described in
\S~\ref{sec:LES}.

For the $32^\circ$ afterbody, the effects of compressibility on the recirculation region and the subsequent formation and evolution of the streamwise vortex pair are examined in \S~\ref{sec:32}.
The vortex cores are identified using the $\Gamma_1$/$\Gamma_2$ method of~\cite{graftieaux2001combining}, allowing their trajectories, cross-sectional areas, and circulation to be tracked through the wake.
%Together, these quantities distinguish changes in the location and size of the vortices from changes in their strength and downstream growth.
The corresponding surface-pressure distributions are also examined to relate the altered vortex structure to its aerodynamic footprint on the upswept base.

For the $45^\circ$ afterbody, vortex-pair and separated-wake states are known to coexist at incompressible speeds~\citep{ranjan2020hysteresis,ranjan2020mean,zigunov2022hysteretic}. 
Complementary increasing- and decreasing-Mach-number sequences are therefore used to determine whether compressibility can induce a change in wake state and produce hysteresis with Mach number.
The associated changes in flow topology and base pressure, together with the sensitivity of the separated-wake state to Reynolds number, are examined in \S~\ref{sec:45}.
%Because the Mach-number sequences are performed at fixed $Re_D$, the sensitivity of the principal observations to Reynolds number is subsequently assessed in \S~\ref{sec:Re}.
Concluding remarks are made in $\S$~\ref{sec:Conclusion}

\section{\label{sec:LES} Large Eddy Simulation Database}

%The baseline simulations 
%The LES databases are categorized as Baseline

%

\subsection{\label{sec:LES1} Baseline Simulations}
%As stated previously, all the LES simulations presented here are based on the previous simulations of Ranjan et al.~\cite{ranjan2020mean} and Ranjan and Gaitonde~\cite{ranjan2020hysteresis}.
%The 3D compressible Navier-Stokes (NS) equations are solved in generalized curvilinear coordinates using an in-house finite difference flow solver. %their non-dimensional form using an in-house finite
%The NS equations are non-dimensionalized using the 
%The baseline LES simulations correspond to a Mach~0.1 flow %over a $32^\circ$ and $45^\circ$ upsw
The baseline LES database %used in the present study 
corresponds to a Mach~0.1 flow %at $Re_D=2.5 \times 10^4$ 
over the canonical configuration shown in Fig.~\ref{fig:FullGeom} at different $\phi$ values.
The baseline simulations are well validated with experimental data and have been used extensively in several prior publications to understand the flow topology at different $\phi$ values~\citep{ranjan2020mean}, to examine the hysteric behavior with $\phi$ and $Re$~\citep{ranjan2020hysteresis} and to analyze the instability mechanisms resulting in vortex meandering~\citep{ranjan2022instability}. 
%the flow dynamics have been discussed in several prior publications. 
%Examples include understanding the flow topology at different $\phi$ values~\cite{ranjan2020mean}, the examination of hysteric behavior with $\phi$ and $Re$~\cite{ranjan2020hysteresis}, and the analysis of instability mechanisms resulting in meandering~\cite{ranjan2022instability}. 
In this study, we focus our attention to two upsweep angles, $\phi_1=32^\circ$ and $\phi_2=45^\circ$; %are selected for further analysis; 
these test cases are referred to as 32M01 and 45M01V throughout this work.

%\subsection{LES Methodology}
The simulations were performed %presented in this work are performed 
by solving the non-dimensional 3D compressible Navier-Stokes (NS) equations in generalized curvilinear coordinates using an in-house finite difference solver.
The NS equations are non-dimensionalized using the cylinder diameter ($D$), freestream velocity ($U_{\infty}$) and freestream density ($\rho_{\infty}$). 
The freestream pressure is then given by $1/(\gamma M^2)$, where $M$ is the Mach number and $\gamma=1.4$.
The $Re_D$ value is fixed at $2.5 \times 10^4$.
Although this $Re_D$ value is significantly lower than actual flight conditions, previous simulations~\citep{ranjan2020mean} and experiments~\citep{bulathsinghala2017afterbody,epstein1994experimental,zigunov2019flow} have shown that the key features of interest are relatively independent of $Re_D$.
Thus, $Re_D=2.5 \times 10^4$ reflects a balance between resolving flow features of interest and the computational cost.  

The spatial discretization is performed using a fourth-order compact scheme~\citep{lele1992compact,gaitonde1998high} with a sixth-order filter~\citep{gaitonde2000pade}. 
The filter ensures numerical stability by damping out the under-resolved high-frequency fluctuations, and provides an implicit subgrid closure mechanism for the LES~\citep{visbal2002large,garmann2013comparative}.
Time integration is achieved using a second-order implicit diagonalized~\citep{pulliam1981diagonal} Beam-Warming scheme~\citep{beam1978implicit}.
The solution is marched forward in time using a constant non-dimensional time step of $\Delta t=2.5 \times 10^{-4}$. 
This $\Delta t$ is found sufficient to accurately resolve the wake of the upswept afterbody %growth of the downstream vortices 
while ensuring temporal accuracy.

\begin{figure}
    \centering
    \includegraphics[width=\textwidth]{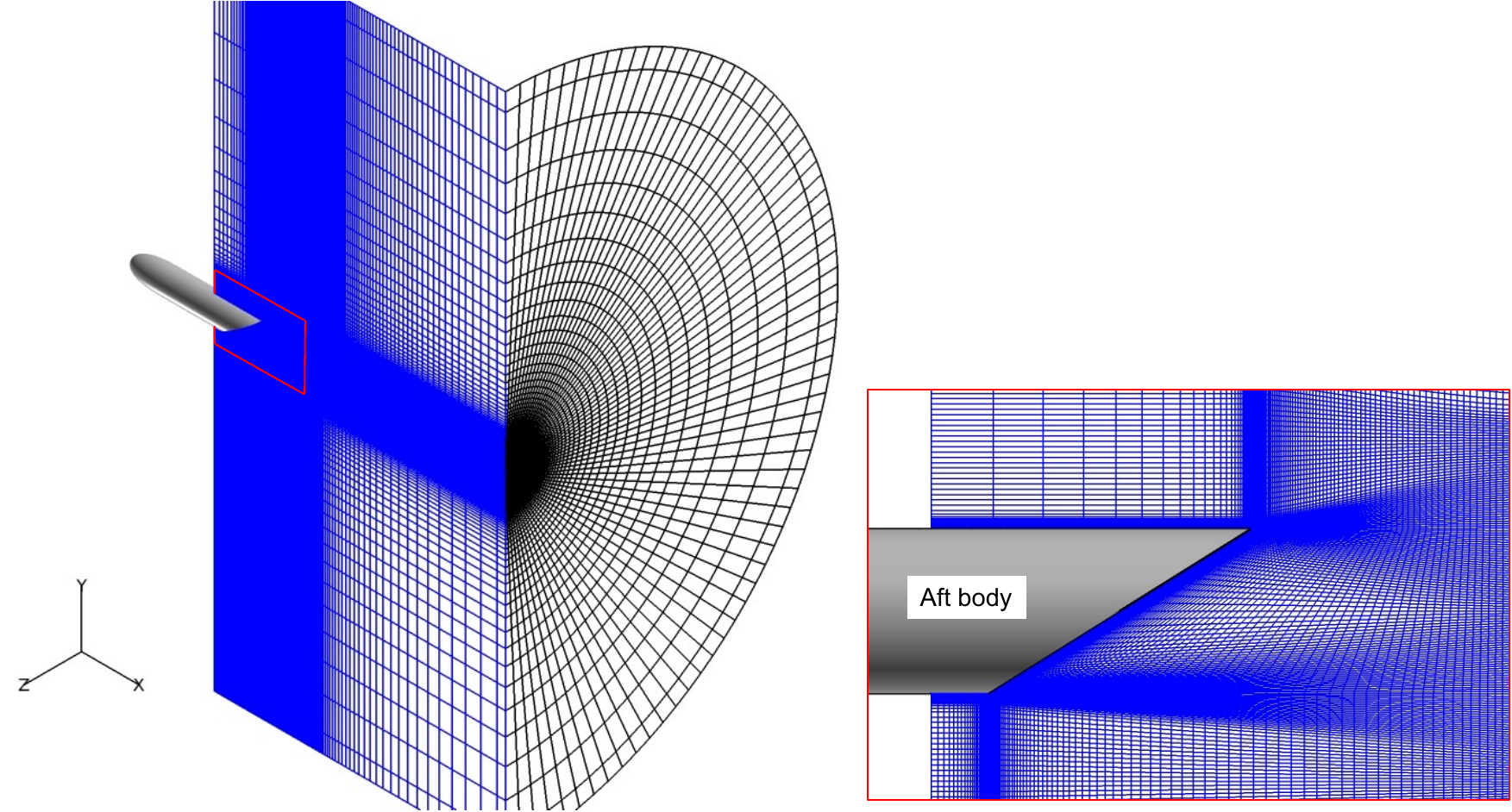}
    \caption{A schematic of the computational domain for the upswept cylindrical aftbody..}
    \label{fig:schematic}
\end{figure}

%\subsubsection{32M01}
%We first focus on the 32M01 case. 
Figure~\ref{fig:schematic} shows a schematic of half of the computational domain for the 32M01 case.
%The mesh for the 45M01V case follows the same topology.
The computational domain extends $12D$ downstream and $14D$ in the radial directions.
The grid consists of $498$, $418$ and $285$ points in the streamwise ($x$), radial ($r$) and azimuthal ($\theta$) directions respectively; every fourth mesh point along a representative streamwise and cross plane is shown in Fig.~\ref{fig:schematic} for completeness.
The mesh is clustered around the aft-body geometry with gradual stretching along the outer boundaries.
%Additional details of the grid topology can be found in Ranjan et al.~\cite{ranjan2020mean}.
The mesh density in the clustered regions is comparable to that of %Garman and Visbal~
\cite{garmann2019high}, who performed numerical simulations for the baseline variant at much higher $Re_D$ values.

The origin is placed at the center point of the straight line joining the upstream and downstream apexes.
The incoming flow is aligned along the $x-$ axis as shown.
%Typically, experiments performed on consist of an ellipsoidal
%For computational efficiency, 
To reduce the computational complexity, the simulation domain does not include the ellipsoidal forebody %that is typically present in the canonical configuration 
shown in Fig.~\ref{fig:FullGeom}. %, is not simulated. 
Instead, the inflow consists of an experimentally matched boundary layer profile~\citep{ranjan2020mean} from a precursor axisymmetric simulation at the same $Re_D$.
No-slip, adiabatic boundary conditions are applied at the solid surface of the fuselage, whereas all the radial boundaries are considered as freestream. 
Since the downstream boundary is sufficiently far away, a zero-gradient condition is imposed at this boundary.
These boundary conditions have been proven to correctly replicate the experimentally observed vortex dynamics at these operating conditions~\citep{ranjan2020mean}.
The mesh consists of a singularity along the $x-$axis; this singularity is treated as a boundary condition by enforcing solution continuity in the manner described in %Gaitonde
~\cite{gaitonde2012analysis}.
Additional details of the grid topology, boundary conditions and mesh independence studies can be found in %Ranjan et al.
~\cite{ranjan2020mean}.
%The solution approach is identical to the one employed by Garman and Visbal~\cite{} for much higher Reynolds numbers.

\subsection{\label{sec:Flowstruc1} Overall Flow Structure}

\begin{figure}
    \centering
    \subfloat[]{\includegraphics[width=0.5\textwidth]{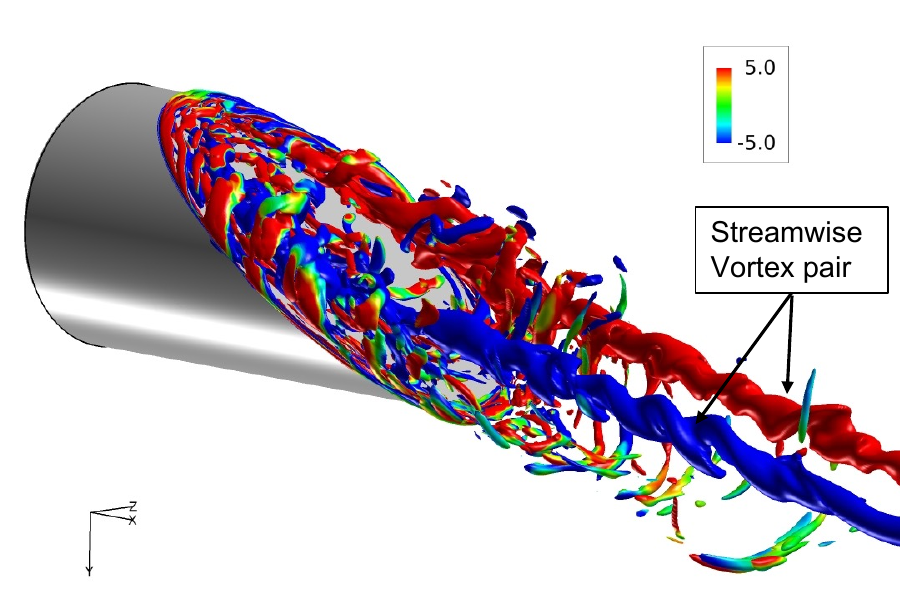}} 
    \subfloat[]{\includegraphics[width=0.5\textwidth]{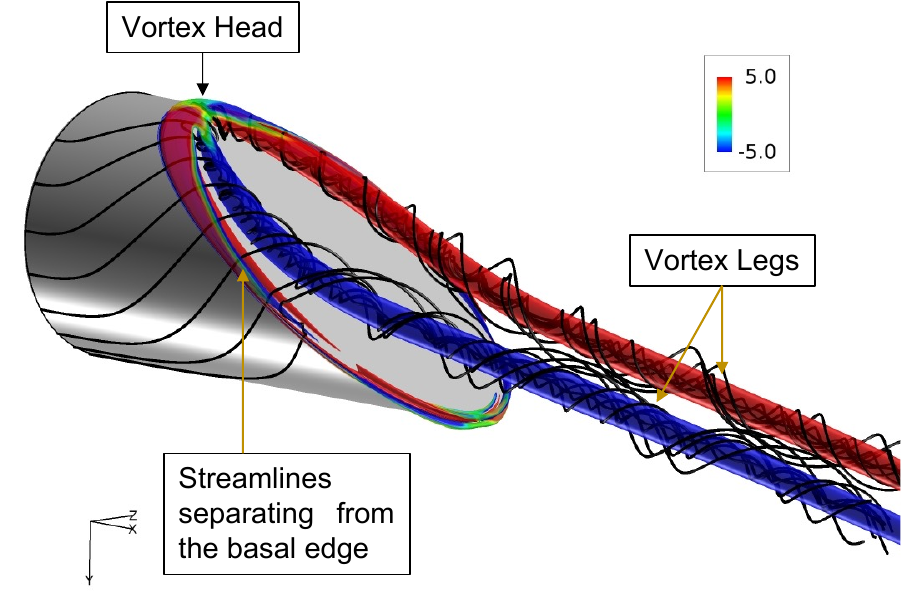}}
    %\subfloat[]{\includegraphics[width=0.5\textwidth]{Compressibility_Effects/Figures/Baseline32_Inst.pdf}} 
    %\subfloat[]{\includegraphics[width=0.5\textwidth]{Compressibility_Effects/Figures/Baseline45_Inst.pdf}}
    \caption{32M01: (a) Instantaneous and (b) Mean $Q$-criterion iso-surfaces ($Q=37$) colored with streamwise vorticity. Three-dimensional streamlines in (b) illustrate the roll-up of flow separating from the base edge into the vortex head and streamwise legs.}
    \label{fig:32M01}
\end{figure}
Figure~\ref{fig:32M01}\textcolor{red}{a} shows the instantaneous $Q-$criterion iso-surfaces ($Q=37$) colored with streamwise vorticity for the 32M01 case at an arbitrary time-step. 
%The $Q-$criterion iso-surface highlights the chaotic nature of the flow over the flat base, with interactions between different scales. 
The $Q-$criterion iso-surface highlights the complex and chaotic nature of the flow over the flat base, characterized by interactions between different scales.
A streamwise-oriented vortex pair is clearly visible, emerging from these interactions. However, a clearer picture of the vortex pair can be obtained by observing the mean flow as shown in Fig.~\ref{fig:32M01}\textcolor{red}{b}.
The mean flow is obtained by averaging the unsteady LES snapshots over approximately $200$ characteristic times based on $D$ and $U_\infty$; this duration is found sufficient to provide a converged mean flow~\citep{ranjan2020mean}. %providing a converged mean flow.
%In order to clearly he
%The instantaneous $Q-$criterion snapshot highli
For reference, three-dimensional streamlines are superimposed over the mean $Q-$criterion iso-surface. 
Unlike the instantaneous flow, the mean $Q-$criterion iso-surface displays a well-defined horseshoe-like vortex structure with a prominent vortex head and legs that orient themselves to form a counter-rotating vortex pair. 
The vortex head consists of recirculating streamlines near the upstream apex. 
The vortex grows over the upswept base, entraining fluid from the shear layer separating along the periphery of the flat base. 
Ultimately, the vortex lifts off from the upswept surface to form the counter-rotating streamwise vortex pair.
%\textbf{Cite other people about details}

\begin{figure}
    \centering
    \subfloat[]{\includegraphics[width=0.5\textwidth]{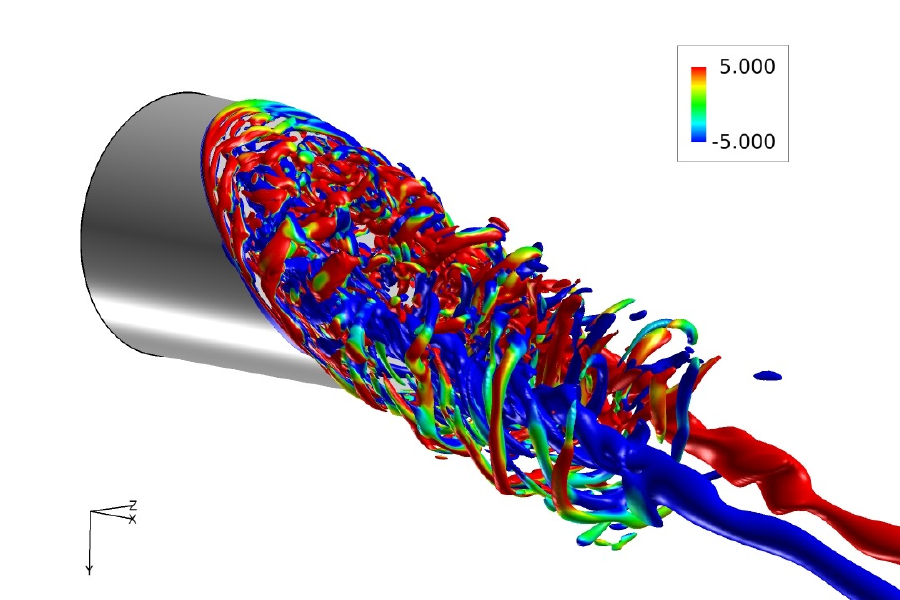}} 
    \subfloat[]{\includegraphics[width=0.5\textwidth]{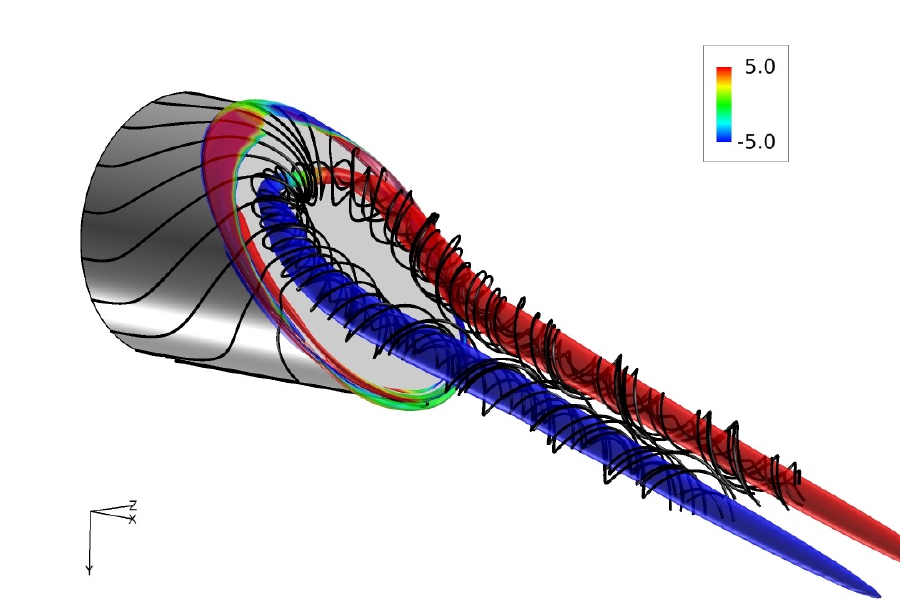}}
    %\subfloat[]{\includegraphics[width=0.5\textwidth]{Compressibility_Effects/Figures/Baseline32_Inst.pdf}} 
    %\subfloat[]{\includegraphics[width=0.5\textwidth]{Compressibility_Effects/Figures/Baseline45_Inst.pdf}}
    \caption{45M01V: (a) Instantaneous and (b) Mean $Q$-criterion iso-surfaces ($Q=37$) with 3D streamlines.}
    \label{fig:45M01V}
\end{figure}

Figure~\ref{fig:45M01V} shows the instantaneous and mean $Q-$criterion contours ($Q=37$) for the 45M01V test case.
Compared to the 32M01 case, the 45M01V case exhibits three major differences.
First, the unsteady flow appears more chaotic and the vortex pair is not so easily discernable since it is surrounded by small scales that interact vigorously with the large vortices. 
Second, similar to the 32M01 case, the 45M01V mean flow also exhibits a horseshoe type vortex.  %with certain noteceable differences.
However, the vortex head is formed further downstream compared to 32M01; this downstream shift in the vortex head can be attributed to the increase in the separation bubble size at the upstream apex. 
Further ramifications of this larger recirculation region are discussed later in $\S$~\ref{sec:IV}.
%As a result, the vortex head is formed further downstream. 
Finally, the vortex legs are much larger in diameter for the 45M01V case.
The larger size of these vortices implies a larger drag penalty for the $45^\circ$ case %on the flat base 
as reported previously both experimentally~\citep{bulathsinghala2017afterbody,zigunov2020reynolds} and numerically~\citep{ranjan2020mean}.
A more comprehensive description of the differences between the two cases can be found in %Ranjan et al.~
\cite{ranjan2020mean}.
%Effect of angle: trajectory, size of vortices. More details in Ranjan et al.
%\begin{figure}
%    \centering
%    \subfloat[]{\includegraphics[width=0.5\textwidth]{Compressibility_Effects/Figures/Baseline32_Mean.pdf}} 
%    \subfloat[]{\includegraphics[width=0.5\textwidth]{Compressibility_Effects/Figures/Baseline45_Mean.pdf}}
%    \caption{Mean $Q$-criterion iso-surfaces ($Q=37$) for the baseline simulations at $\phi=32^\circ$ (top) and $\phi=45^\circ$ (bottom).}
%    \label{fig:BaselineSimMean}
%\end{figure}
%Figure~\ref{fig:BaselineSimUnsteady}a shows blah blah. 

\subsection{\label{sec:LES2} Simulation Sequences to Examine Compressibility Effects}
\begin{figure}
    \centering
    \includegraphics[width=\textwidth]{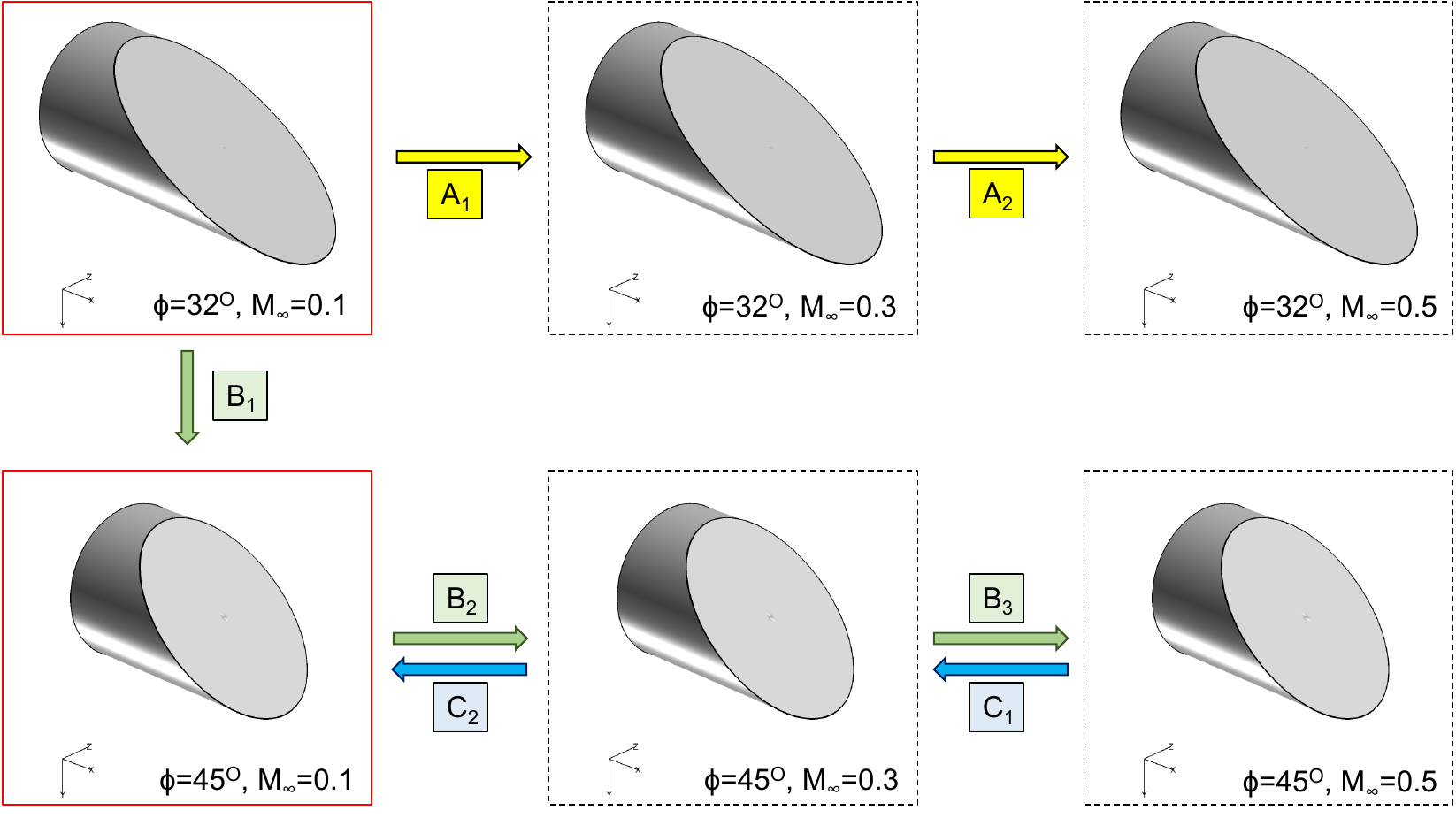}
    \caption{Simulation sequences to assess the effect of Mach number of the aftbody wake. The baseline simuations taken from \citet{ranjan2020mean} and \citet{ranjan2020hysteresis} are placed inside red boxes.}
    \label{fig:SimSeq}
\end{figure}

%In order to examine the effect of compressibility on the flow-field, three simulation sequences are performed.
%In order to investigate the effect of compressibility on the flow-field, three simulation sequences are performed.
We examine the effect of compressibility on the flow-field %is examined 
through three carefully designed simulation sequences.
%The $Re_D$ value is kept constant throughout these sequences; this allows 
Figure~\ref{fig:SimSeq} depicts these three sequences. 
%Table~\ref{tab:allcases} provides a list of all the resulting test cases from these sequences.
%A list of all the test cases is documented in Table~\ref{tab:allcases}.
\begin{table}
    \centering
    \caption{List of all test cases used in this investigation.}
    %\begin{tabular}{m{1cm}| m{2cm} m{1cm} m{2cm} m{2cm} m{3cm}|}
    \begin{tabular}{c c c c c c c}
    %\hline
       S.No & \hspace{0.5cm} Test Case \hspace{0.5cm} & \hspace{0.1cm}$\phi$\hspace{0.1cm} & \hspace{0.2cm}Mach No.\hspace{0.2cm} & \hspace{0.1cm}$Re_D$ \hspace{0.1cm}& \hspace{1cm} Regime \hspace{1cm} & Precusor Simulation \\ \hline
        1 & 32M01 & $32^\circ$ & 0.1 & $2.5 \times 10^4$ & Vortex-pair & Baseline\\
        2 & 32M03 & $32^\circ$ & 0.3 & $2.5 \times 10^4$ & Vortex-pair & 32M01\\
       3 & 32M05 & $32^\circ$ & 0.5 & $2.5 \times 10^4$ & Vortex-pair & 32M03\\
        4 & 45M01V & $45^\circ$ & 0.1 & $2.5 \times 10^4$ & Vortex-pair & 32M01\\
        5 & 45M03V & $45^\circ$ & 0.3 & $2.5 \times 10^4$ & Vortex-pair & 45M01V\\
        6 & 45M05W & $45^\circ$ & 0.5 & $2.5 \times 10^4$ & Separated-Wake & 45M03V \\
        7 & 45M03W & $45^\circ$ & 0.3 & $2.5 \times 10^4$ & Separated-Wake & 45M05W\\
        8 & 45M01W & $45^\circ$ & 0.1 & $2.5 \times 10^4$ & Separated-Wake & 45M03W\\ %\hline
        9 & 45M03W-75K & $45^\circ$ & 0.3 & $7.5 \times 10^4$ & Separated-Wake & 45M03W\\
   \end{tabular}
    
    \label{tab:allcases}
\end{table}
%The first 
The simulation sequences %in Fig.~\ref{fig:SimSeq} 
are labeled from A through C. 
Each sequence consists of multiple steps.
For instance, sequence A is comprised of two steps A$_1$ and A$_2$, and focuses on the $\phi=32^\circ$ geometry.
Beginning with the 32M01 case, the freestream Mach number is increased to $0.3$ while maintaining a constant inflow boundary layer height ($\delta$) and $Re_D$ (step A$_1$).
%The first sequence (A1-A2) is performed on the $\phi=32^\circ$ geometry.
%Starting from the 32M01 case, the Mach number is increased $0.3$ while keeping the inflow boundary layer height ($\delta$) and $Re_D$ constant.
%The constant $Re_D$ allows a straightforward m
The simulation is run for over $400$ convective time units to ensure statistical stationarity. 
%Care is taken to ensure that it is run for enough time steps.
Once stationarity is achieved, the Mach number is further increased to $0.5$ while keeping $\delta$ and $Re_D$ constant (step A$_2$). 
%Once the solution has reached statistical stationarity, the Mach number is again increased to $0.5$ keeping $\delta$ and $Re_D$ constant.
%Keeping $\delta$ and $Re_D$ constant allows us to dilineate the effect of Mach number only.
%This assumes that the flow features are independent of $Re_D$. This assumption is tested later in section x.
This approach isolates the effects of Mach number, assuming that the flow characteristics are independent of $Re_D$; previous investigations~\citep{ranjan2020mean} have shown that the main topological features of the flow are relatively independent of $Re_D$ at this $\phi$.
The resulting new cases are labelled as 32M03 (Mach~$0.3$) and 32M05 (Mach~$0.5$).

Sequence B comprises of three steps, B$_1$ through B$_3$ and follows the same procedure as sequence A. 
The goal of sequence B is to investigate the role of compressibility on the $\phi=45^\circ$ configuration. 
The 32M01 flow-field is used as an initial condition for the 45M01V case. 
This choice is based on previous observations~\citep{zigunov2020reynolds,ranjan2020hysteresis} that show the co-existence of a `vortex-pair' state and a `separated-wake' state at $\phi=45^\circ$. 
Thus, using the 32M01 flow-field as an initial condition ensures the persistence of the vortex-pair for 45M01V.
This approach has been previously tested by \cite{ranjan2020hysteresis} to examine the hysteric behavior with $\phi$ and \textit{Re} at baseline operating conditons.
Starting from the 45M01V case, which was derived from the 32M01 case, the freestream Mach number is increased to $0.3$ and subsequently to $0.5$, resulting in cases 45M03V and 45M05W respectively.
%The new cases are named 45M03V and 45M05W for later use. 
As demonstrated later in $\S$~\ref{sec:IV}, the increase in Mach number from $0.3$ to $0.5$ causes a transition from a `vortex-pair' state to a `separated-wake' closure.

The final simulation sequence (C) is designed to investigate the occurrence of hysteresis with Mach number.
The sequence begins with the separated-wake 45M05W state. 
The freestream  Mach number is then reduced to $0.3$ (step C$_1$) and subsequently to $0.1$ (step C$_2$), resulting in cases 45M03W and 45M01W, respectively. 
The objective is to determine if the flow regime exhibits a hysteresis effect, where the separated-wake state persists even when the Mach number is decreased back to $0.1$. 
The results from sequence C are discussed in $\S$~\ref{sec:Hysteresis}.

Table~\ref{tab:allcases} provides a list of all the resulting test cases from these sequences. Nine cases are necessary, three for  $32^o$ and four for $45^o$ upsweep. The different Mach numbers are listed in the fourth column.  
The Reynolds number is kept constant, except for the last case, while the last column lists the initial condition for each simulaton.

%\begin{figure}
%    \centering
%    \subfloat[]{\includegraphics[width=0.5\textwidth,trim=0 0 350 0,clip]{Compressibility_Effects/Figures/M01_InstQ37_3D.jpeg}} \\
%      \subfloat[]{\includegraphics[width=0.7\textwidth,]{Compressibility_Effects/Figures/M01_InstQ37_ZView.jpeg}}
%    \caption{Caption}
%    \label{fig:my_label}
%\end{figure}

% DVG: Somewhere put that for sharp edge, Re is not that important.
\section{Sequence A: Flow over ${\textbf{32}^\circ}$ afterbody} \label{sec:32}
%\section{Result}
\subsection{Recirculation Zone}
\begin{figure}
    \centering
 %\subfloat[]{\includegraphics[width=0.65\textwidth,trim=0 0 300 0, clip]{Compressibility_Effects/Figures/32M03_MeanQ37_newview.jpeg}}\\
  %\subfloat[]{\includegraphics[width=0.65\textwidth,,trim=0 0 300 0, clip]{Compressibility_Effects/Figures/32M05_MeanQ37_newview.jpeg}}
 %\subfloat[]{\includegraphics[width=0.5\textwidth]{Compressibility_Effects/Figures/45MeanQ_YView.pdf}}
\includegraphics[width=\textwidth]{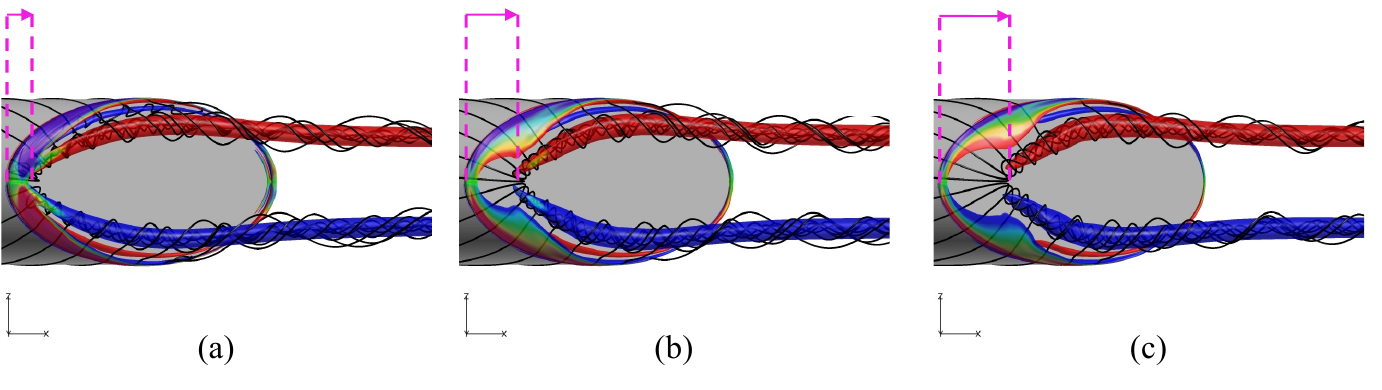}
    \caption{Sequence A: Mean $Q-$criterion iso-surfaces for (a) 32M01, (b) 32M03 and (c) 32M05.}
    \label{fig:32MeanQ}
\end{figure}

We first examine the results of sequence~A. 
Figure~\ref{fig:32MeanQ} shows the mean $Q-$criterion plots superimposed with 3D streamlines for all the $32^\circ$ cases. 
For reference, dashed vertical lines identify the streamwise locations of the upstream apex and the vortex head, while the solid arrow denotes the distance between them.
%For reference, the streamwise distance between the upstream apex and the vortex head are highlighted using dashed vertical lines.
%The streamwise distance between the vortex head and the upstream apex is marked using a solid arrow.
The main features of the flow-field are qualitatively similar at all Mach numbers at this upsweep angle, with the horeshoe vortex described in $\S$~\ref{sec:LES1} persisting for all cases.
%The horseshoe vortex comprising of a vortex head 
%\textbf{Need to build some richness here}.
However, significant differences are quite prominent near the upstream apex.
%It is observed that 
The formation of the vortex head is pushed further downstream with increasing Mach number.
%Since the vortex head is directly related to the flow separation at the upstream apex, this observation warrants a closer look at the flow-field %near the upstream apex. %
%on the $z=0$ symmetry plane.
Because the vortex head develops from the separated flow near this
apex, its downstream displacement suggests a corresponding change in
the size and topology of the recirculation region.
%%CP: Add comparisons with backward facing step or recirculation increasing with Mach number.
% DVG: Read carefully and make sure F1 and F2 are in all described

\begin{figure}
    \centering
    \subfloat[]{\includegraphics[width=0.85\textwidth]{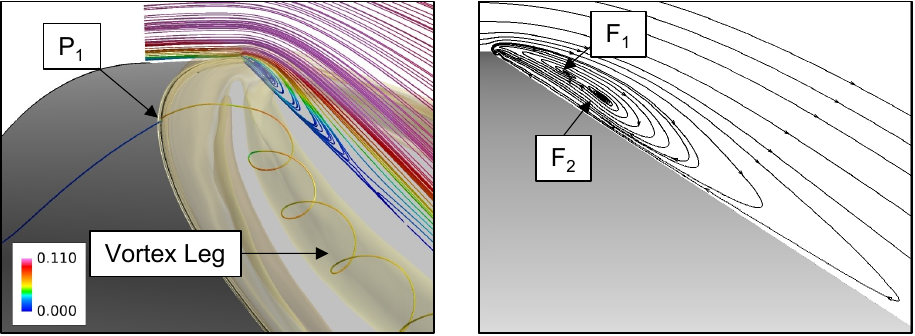}} \\
    \subfloat[]{\includegraphics[width=0.85\textwidth]{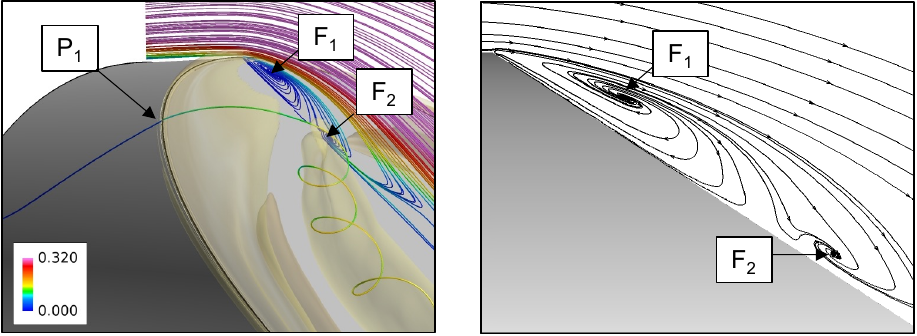}} \\
    \subfloat[]{\includegraphics[width=0.85\textwidth]{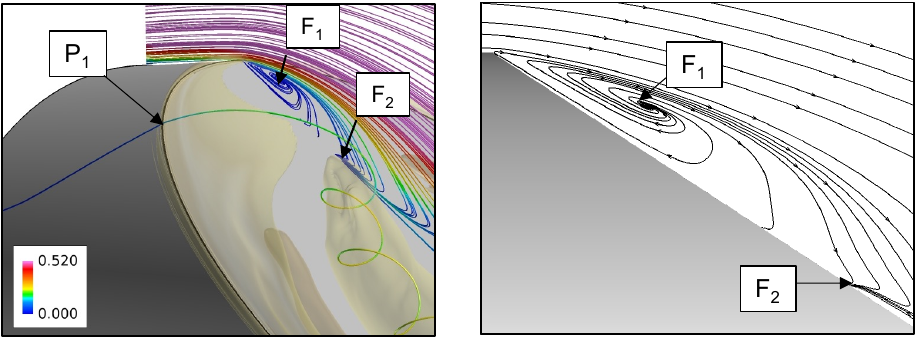}}
    \caption{Sequence A: Flow near the upstream apex for (a) 32M01, (b) 32M03 and (c) 32M05.}
    \label{fig:32SepBubble}
\end{figure}

%\subsection{Symmetry Plane}
%The flow near the upstream apex is examined in Fig.~\ref{fig:32SepBubble} using streamlines along the $z=0$ symmetry plane.
Figure~\ref{fig:32SepBubble} examines this region using streamlines on
the $z=0$ symmetry plane.
%shows the 2D streamlines colored with Mach number on the the $z=0$ symmetry plane.
%The streamlines are colored with Mach number to highlight the separation bubble (denoted as SB in Fig.~\ref{fig:32SepBubble}a).
The streamlines are colored by Mach number, with contour levels selected separately for each flow field to expose the local structure.
%Note that Mach contours are plotted at different levels based on %local maximum and minimum values in 
%the respective flow-fields. 
%For reference, the horseshoe vortex is also highlighted for all the three cases.
The corresponding horseshoe vortex and a representative streamline passing through the azimuthally offset point P$_1$ are also shown for reference.
At each Mach number, the symmetry-plane streamline portrait contains two critical focii, denoted by F$_1$ and F$_2$.
F$_1$ is a source-type focus (streamlines move away from the point) located near the sharp upstream edge, whereas F$_2$ is a sink-type focus (streamlines moving toward the point) within the recirculating flow.
Streamlines emerging from the vicinity of F$_1$ wrap around the separated region and spiral toward F$_2$.
The latter represents the symmetry-plane trace of the vortex head identified in Fig.~\ref{fig:32MeanQ}.

%For the 32M01 case, the trace of the separation bubble in the symmetry plane consists of a re-circulation region with a singular node, N. %; this node coincides with the vortex head. 
%The node N represents the trace of the vortex head on the symmetry plane.
%As the Mach number increases, the size of the separation bubble increases.
%At higher Mach numbers, the recirculation region in the symmetry plane consists 
%With an increase in Mach number, the size of the separation bubble increases, and the symmetry plane now consists of two nodes (marked as N$_1$ and N$_2$ in Figs.~\ref{fig:32SepBubble}b and~\ref{fig:32SepBubble}c).
%\textbf{Need some stuff about topology of flow separation}.
%The node N$_2$ represents the vortex head in the symmetry plane for both 32M03 and 32M05 cases.
%Moreover, for the 32M05 case, the node N$_2$ is formed further downstream when compared to 32M03.
%For reference, a streamline passing through an azimuthally offset point P1 is also shown for all cases.
%Because of the sharp edge of the base, the onset of separation is fixed by the geometry.
%However, the reattachment point point moves downstream on the base as the flow becomes more compressible.
The location of F$_1$ remains nearly unchanged as the Mach number increases because the sharp edge geometrically fixes the onset of separation.
In contrast, F$_2$ moves progressively downstream, producing an elongation of the recirculation region between the upstream apex and the vortex head.
The downstream displacement is also evident from the streamline passing through P$_1$, which reattaches farther along the upswept base as the Mach number increases.
%This elongation of the recirculation zone with
%Mach number is unsurprising, as it %relates to the pressure gradient and 
A similar trend has been observed in several flows including 
%Examples include 
flow separation over cylinders~\citep{canuto2015two}, triangular airfoils~\citep{suwa2012compressibility}, delta wings~\citep{luckring2002reynolds} and axisymmetric afterbodies~\citep{presz1976analytical}.
%For example, %Suwa et al.~
%\cite{suwa2012compressibility} found that increasing Mach number from $0.15$ to $0.7$ delays reattachment of the separated shear layer on a triangular airfoil. 
%Similarly, for flow past a cylinder, this effect manifests as an increase in the wake size as observed by %Canuto and Taira~
%\cite{canuto2015two}. 
%For an axisymmetric afterbody with a circular arc base, %Presz Jr and Pitkin~
%\cite{presz1976analytical} experimentally noted similar movement of separation and reattachment points as the Mach number increased from $0.25$ to $0.7$. 
%Likewise, on a $65^\circ$ delta wing with a blunt edge, %Luckring~
%\cite{luckring2002reynolds} showed that leading edge separation is significantly promoted when Mach number is increased from $0.4$ to $0.6$.
%This effect can also be seen at other azimuthal planes in the present case. 
%For reference, a streamline passing through an azimuthally offset point P$_1$ is also shown in Fig.~\ref{fig:32SepBubble}.
%Because of the sharp edge of the base, the onset of separation is fixed by geometry for all the three cases.
%However, the reattachment point moves downstream on the base as the flow becomes more compressible.
%\textbf{feeds the vortex}

Because the onset of separation is fixed by the sharp basal edge, the principal characteristics of the separated flow are relatively insensitive to $Re_D$.
Comparisons between simulations and experiments spanning substantially different $Re_D$ values have shown that the primary mean-flow features remain qualitatively similar, including the separation bubble near the upstream apex, its development into a horseshoe-shaped vortex, and the subsequent formation of the streamwise vortex pair~\citep{ranjan2020mean}.
Moreover, since $Re_D$ is held constant in sequence A, the progressive downstream movement of F$_2$ and the associated elongation of the recirculation region can therefore be attributed to the increase in Mach number.
The separating shear layer subsequently rolls up and feeds the legs of the horseshoe vortex, so this delayed reattachment also shifts the initial formation of the streamwise vortex pair downstream.

\subsection{Vortex Evolution}
The separated shear layer originating at the basal edge supplies the vorticity from which the horseshoe-vortex head and its streamwise legs develop.
Because increasing Mach number displaces the reattachment region and the vortex head downstream, it is expected to alter not only the inception of the vortex pair but also its subsequent growth and trajectory.
%The state of the separated layer % near the upstream apex 
%is a major factor in the formation of the downstream streamwise vortex pair as shown previously in Fig.~\ref{fig:32M01}.  
%Moreover, since the inception of these vortices nearly coincides with the reattachment point, it is expected that the vortex structure as well as its strength will change with compressibility.
%Figure~\ref{fig:32xslices} shows the streamwise vorticity at 5 axial stations along the base.

\begin{figure}
    \centering
     \subfloat[]{\includegraphics[width=\textwidth]{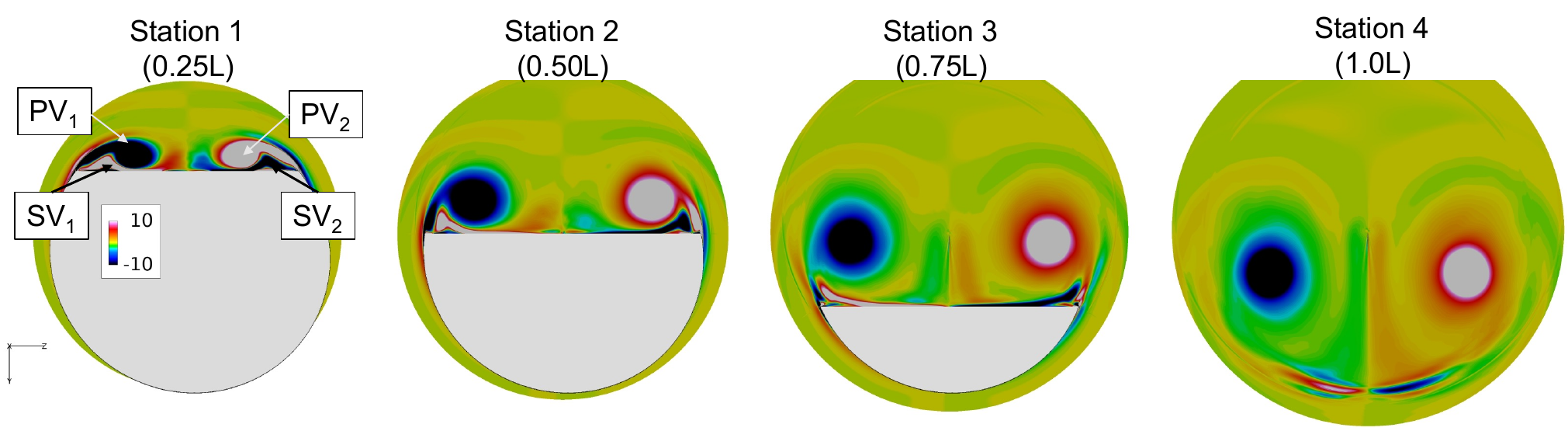}}\\
    \subfloat[]{\includegraphics[width=\textwidth]{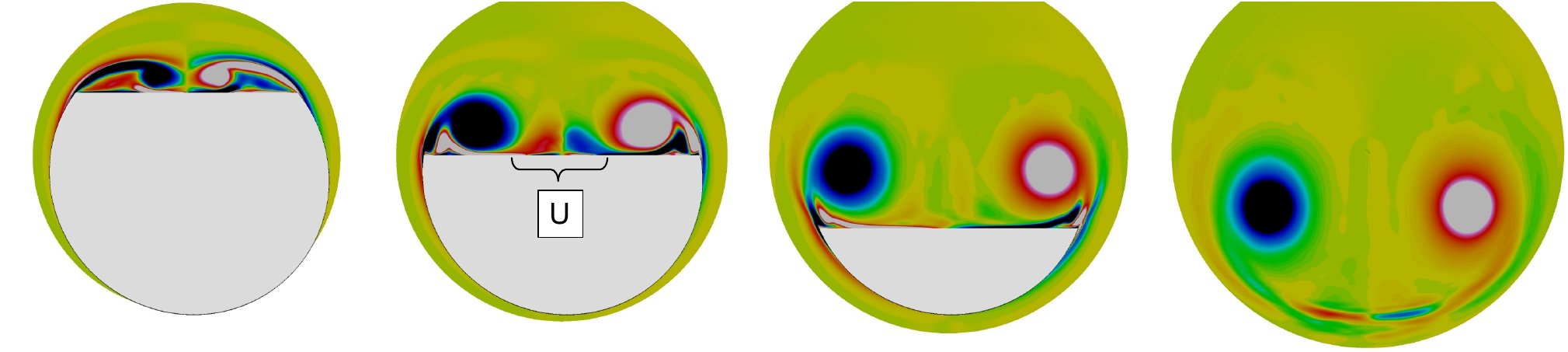}}\\
    \subfloat[]{\includegraphics[width=\textwidth]{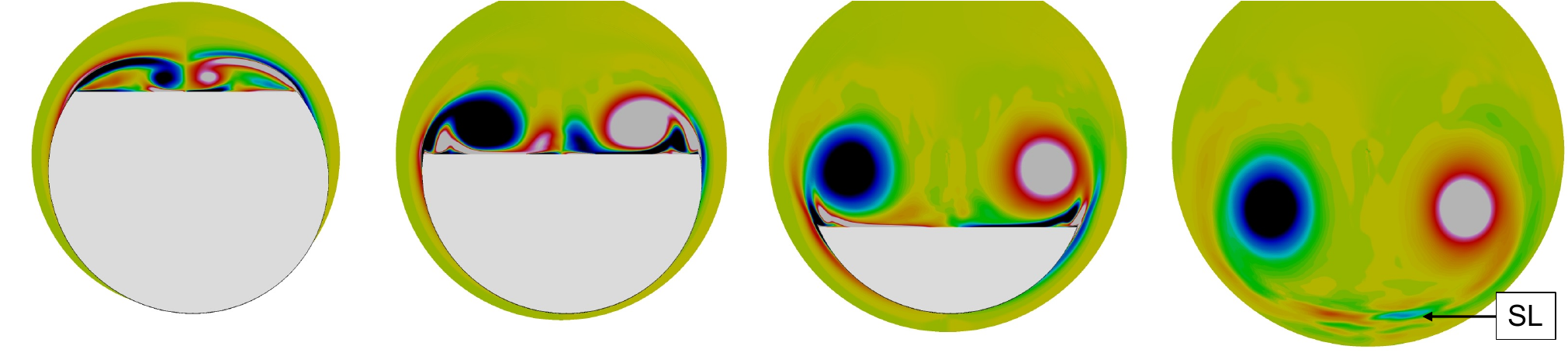}}
     % \subfloat[]{\includegraphics[width=0.55\textwidth]{Compressibility_Effects/Figures/32M03_xslices.pdf}} \\
      %  \subfloat[]{\includegraphics[width=0.55\textwidth]{Compressibility_Effects/Figures/32M05_xslices.pdf}}
    \caption{Evolution of streamwise vorticity at four axial stations for (a) 32M01, (b) 32M03 and (c) 32M05.}
    \label{fig:32xslices}
    \end{figure}

\begin{figure}
    \centering
    \includegraphics[width=\textwidth,trim=80 200 80 200,clip]{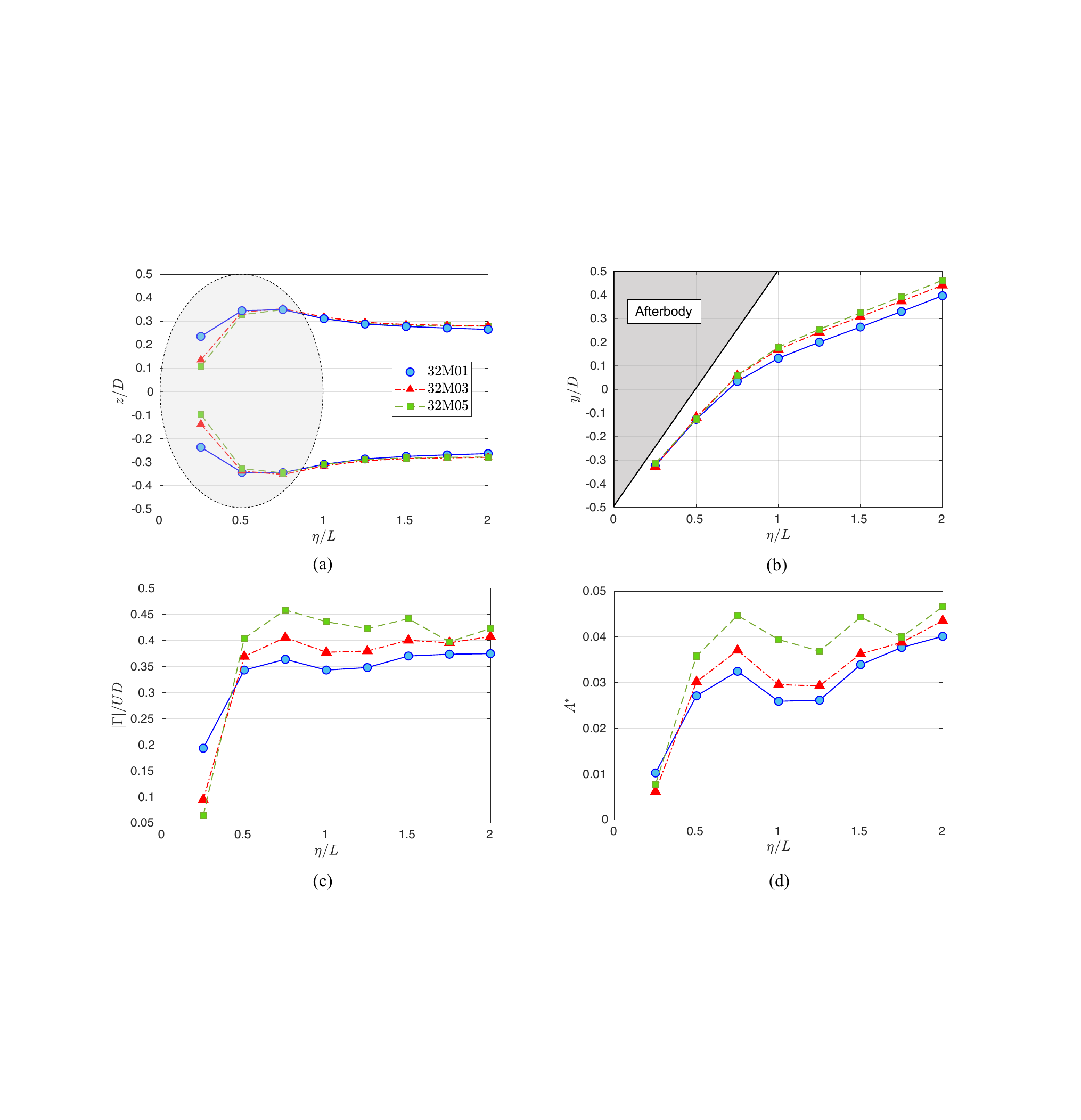}
    \caption{Vortex tracking results for the $32^\circ$ afterbody: (a) $y/D$ vs $\eta/L$, (b) $z/D$ vs $\eta/L$, (c) $|\Gamma|/UD$ vs $\eta/L$ and (d) $A^*$ vs $\eta/L$. Note that the two vortices are symmetrically placed about the $z=0$ plane and therefore $y/D$, $|\Gamma|/UD$ and $A^*$ values for only the left vortex are reported. }
    \label{fig:32graft}
\end{figure}

Figure~\ref{fig:32xslices} compares the mean streamwise vorticity evolution ($\omega_x$) at four axial stations for the three test cases; this views the legs of the horseshoe vortex as two separate vortices. 
%To describe the development of these vortices, five stations are shown along the base and in the wake region where further details of the flow will be presented.  
The stations are designated 1 to 4, and are located at $\eta/L = 0.25, 0.5, 0.75$ and $1.0$, where $L=D \cot{\phi}$ is the distance between the upstream and the downstream apexes and $\eta= x - L/2$
%the axial length of the base, is chosen for non-dimensionalization; 
is the new axial co-ordinate with its origin at the upstream apex. 
Thus, station~2 is located at half the distance between the two apexes and station~4 represents the streamwise plane containing the downstream apex.
A more quantitative comparison of the vortex pair evolution is obtained using the $\Gamma_1/\Gamma_2$ approach of~\cite{graftieaux2001combining}. %is used for vortex core identification and further characterization. 
%\textbf{Something about the approach and how it can be used for vortex core identification and further characterization.}
The $\Gamma_1$ function identifies the vortex-center location from the local organization of the velocity field, whereas $\Gamma_2$ is used
to delineate the vortex boundary.
These quantities permit the trajectory, cross-sectional area, and circulation of each vortex to be evaluated consistently.
Figures~\ref{fig:32graft}a and~\ref{fig:32graft}b show the spanwise ($z/D$) and vertical ($y/D$) positions of the vortex cores using the $\Gamma_1/\Gamma_2$ technique. 
The streamwise locations range from $\eta/L=0.25$ to $\eta/L=2.0$, the first four of which are coincident with the stations shown in Fig.~\ref{fig:32xslices}, thus allowing a better characterization of the vortex pair. 
Note that the two vortices are symmetrically located about the $z=0$ plane and therefore $y/D$, values of only the left vortex are reported.
Figures~\ref{fig:32graft}c and \ref{fig:32graft}d show the circulation ($|\Gamma|/UD$) and non-dimensional area ($A^*$) of the left vortex with $\eta/L$. 
The $|\Gamma|/UD$ and $A^*$ values of the right vortex are within $\pm 1\%$ of these values are thus omitted for brevity.

%Based on these contours several observations can be made @@@
Several observations can be made to assimilate the results.
%The stations $x/D=-0.45$ and $x/D=-0.15$ show the development of the  
\begin{enumerate}[leftmargin=*]
\item At station-1, the 32M01 case shows the presence of %development of the vortex cores as 
two teardrop-shaped primary vortices (marked as PV$_1$ and PV$_2$ in Fig.~\ref{fig:32xslices}a) over the upswept surface. %for both the test cases.
The teardrop structure is indicative of the separated shear layer at the sharp edge that rolls up to feed the PVs.
Secondary vortices (marked as SV$_1$ and SV$_2$) are also evident on this plane, beneath the tear-drop structure.
The SVs have opposite signs to their respective PVs since they are formed due to the induced effect of PV$_1$ and PV$_2$ on the base.
%\textbf{Richness by comparison to other secondary vortices.}
This PV-SV structures shares similarities with those observed by~\cite{gursul2005unsteady} and~\cite{gordnier2009computational} in delta wings.
%\textbf{So whats the point??}.
The higher Mach number cases, 32M03 and 32M05 both show a similar teardrop structure at station-1.
%The PVs become sma
However, with increasing Mach number, the separating shear layer that forms the PVs attaches at a farther distance from the edge, consistent with the observations made based on point P$_1$ in Fig.~\ref{fig:32SepBubble}.
As a result, the PVs are formed closer to each other %and are smaller in size 
with increasing Mach number.
At Mach~0.1 the two PVs are separated by a $\Delta z/D \sim 0.588$ at station-1 (Fig.~\ref{fig:32graft}a); this distance reduces to $\sim 0.276$ and $\sim 0.195$ for Mach~0.3 and Mach~0.5 respectively.
Another consequence of this increase in reattachment length, is the $\sim 40 \%$ reduction in the area of PVs between Mach~0.1 to Mach~0.5 at station-1 (Fig.~\ref{fig:32graft}d). 
The smaller PVs at higher Mach numbers are associated with smaller $|\Gamma|/UD$ values ($\sim 0.064$ at Mach~0.5 compared to $\sim 0.193$ at Mach~0.1), indicating a reduction in the strength of the vortices at station-1.
%As a result, the strength of the secondary vortices also diminishes with Mach number.
% DVG: Since the vortices form later, the SV are somewhat weaker at a given station at higher M.  
As a result, the secondary vortices are much weaker at Mach~0.3 and are not 
%The secondary vortices are not 
visible at Mach~0.5 at station-1.
Despite these differences, the PVs are located at the same $y/D$ location irrespective of the Mach number (Fig.~\ref{fig:32graft}b).

\item All $\phi=32^\circ$ cases show a similar teardrop structure at station-2 comprising of PVs and SVs. %and an upwash region (marked as U) in between.
The size of the PVs however is much larger when compared to station-1, not only because of the evolution with distance due to entrainment (Fig.~\ref{fig:32graft}d), but as the results show, also because of a Mach number effect. 
For example, the 32M05 case shows a $\sim 24\%$ increase in $A^*$ over 32M01 case.
This observation is quite remarkable and indicates that the smaller PVs at Mach~0.5 in station-1 grow in strength much faster than the larger PVs at Mach~0.1 due to compressibility. %observation suggests that the small vortices at station-1 
% DVG: Will change this to "We show that..."
%It is postulated that this rapid increase in size and strength with Mach number is associated with the growing contributions of baroclinic torque and dilation terms that affect the vorticity generation~\cite{}.
%\textbf{Connect to dilatation, barocilinic torque}.
Despite the increase in size and strength with Mach number, the PVs are located at the same ($y/D, z/D$) coordinates (Figs.~\ref{fig:32graft}a and~\ref{fig:32graft}b) for all three cases. 
Additionally, an upwash region (marked as U in Fig.~\ref{fig:32xslices}b) %, having opposite sign compared to the PVs, 
also exists between the two teardrop shapes at all Mach numbers.
The primary vorticity component in this region has the same sign as in the secondary vortices, and increases in strength as the Mach number increases, consistent with the
%This increase in strength of U is directly related to the %increase in 
larger $|\Gamma|/UD$ levels associated with the primary vortices at higher Mach numbers at station-2 (Fig.~\ref{fig:32graft}c). 
%{
%This increase in circulation is remarkable considering the \textbf{comment about maximum vorticity levels of PVs, or larger circulation at higher Mach numbers}
%At station 2, the 
%Second, \textbf{upwash} for the baseline case, the teardrop vortices (P1 and P2) observed at station A1 evolve into a more axisymmetric shape downstream (A2 to A4) due to entrainment and diffusion processes. 
%Moreover, these vortices are no longer attached to the upswept base.
%This explains the positive $C_p$ values over the base surface at this axial station as shown previously.
%For the modified geometry cases however, the shape of these vortices is more oval.
%Moreover, the $\omega_x$ levels associated with these oval-shaped vortices (station A4) are much higher than those associated with the swirling motion observed at station A2 for Mod03 and the teardrop vortices at the station A1 for Mod03hiRe. %swirling motion observed at upstream stations.
%This increase in $\omega_x$ levels is consistent with the previous streamline plots (Fig.~\ref{fig:Streamlines}) and indicates that the vortex pair grows in strength much further downstream for the modified geometry.
%}

\item The teardrop vortices observed at station-1 and station-2 evolve into axisymmetric vortices at station 3 due to entrainment and diffusion processes. 
These vortices are no longer attached to the upswept base (as indicated by the absence of the upwash region) irrespective of the Mach number.
%As a result, the upwash region is now absent in station 3.
This observation is especially remarkable, considering the delay in the inception of the vortex head with Mach number as discussed previously.
Despite the late formation of the vortices at station-1 at Mach~0.5, the counter-rotation vortex pair lifts off the surface at roughly the same axial station as Mach~0.1. %; this indicates that the vortex pair grows in strength much faster due to compressibility.
%\textbf{barocilinic torque and dilatation}
%Thus, compressibility indicates that the vortex pair grows in strength much further downstream for the modified geometry.
%Due to the larger separation distance for the baseline case, there exists an upwash region (marked at U on stations A1 and A2) between the two vortices. 
%This upwash region is absent for the modified cases due to the reduced distance between the vortices.
%Additionally, since the vortices are lifted off the surface at station 3, the upwash region is now absent. %at station 3.
The vortex cores are located at the same $z/D$ locations for all three cases and continue to be larger in size and strength at higher Mach numbers. 

\item Station-4 shows the vortex pair leaving the upswept afterbody. 
The shear layer leaving the downstream apex (marked as SL in Fig.~\ref{fig:32xslices}c) is also visible in the $\omega_x$ contours.
This connecting shear layer continues to entrain flow into the vortices and can influence their near-wake motion through vortex--shear-layer interactions~\citep{ranjan2020meandering}.
The shear layer signature is more prominent at lower Mach numbers; this trend is consistent with %is unsurprising as %this observation is consistent with classical experiments performed by
%The lower prominence of the shear layer at higher Mach numbers is uncan be related to the lower growth rate of free shear layers with compressibility%observation is consistent with classical experiments performed by
previous experiments~\citep{bogdanoff1983compressibility,papamoschou1988compressible,elliott1990compressibility} have shown that the growth rate of free shear layers decreases with compressibility. %that can be related to the lower growth rate of 
%\textbf{Effect on cargo dynamics?? Maybe meandering??}.
%Although the vortex pair has lifted from the surface by this location, the downstream-apex shear layer can continue to entrain fluid into the vortices and influence their near-wake evolution~\cite{ranjan2020meandering}.
The reduced prominence at higher Mach numbers therefore suggests weaker residual feeding of the detached vortex pair and an earlier transition toward freely evolving wake vortices.
%Moreover, all test cases show a drop in $|\Gamma|/UD$ and $A^*$ from station-3.
%The vortex pair appears noticeably larger with increasing Mach number and the shear layer is more prominent at lower Mach numbers. 
%\textbf{Drop in circulation at this station.}
%Another aspect related to the vortex development along the base is that area and circulation of the vortex cores at X/L = 1.0 are lower than their values at X/L = 0.6 in both flows. 
%This was earlier observed for several upsweep angles for the baseline flow by Ranjan et al.~\cite{ranjan2020mean}, and indicates a small reduction in vortex strengths once they separate from the base and are no longer fed by the shear layer.

\item The change in Mach number results in a change in trajectory of the vortex cores as shown in Fig.~\ref{fig:32graft}b. 
%at $L$ distance away from the downstream apex, Figure~\ref{fig:32graft}b shows the vertical positions of the cores for the three cases. 
%The $y/D$ variation shows that the vortex cores are formed closer to the flat base with increase in Mach number; this results in an upward shift in the trajectory of the vortex pair. 
This change in trajectory is more noticeable %farther downstream. 
%For instance, 
at $L$ distance away from the downstream apex ($\eta/L=2$), where the 32M05 case shows a shift of $y/D \sim 0.066$ when compared with 32M01.
From a scalability standpoint, since the dynamics of more practical configurations are comprised of much larger length scales ($D$), %while the flow and sound speeds (and hence Mach number) do not change commensurately.
this $\sim 0.066D$ separation distance between Mach~0.1 and Mach~0.5 could have implications on paratrooper safety, payload trajectories and the safe longitudinal separation distance from trailing aircraft. %\textbf{See what Rajesh said when comparing angles in the mean flow paper}
The spanwise distance between the vortex pair however, is independent of Mach number.
\end{enumerate}

\subsection{Base Surface Pressure}
\begin{figure}
    \centering
    \includegraphics[width=\textwidth]{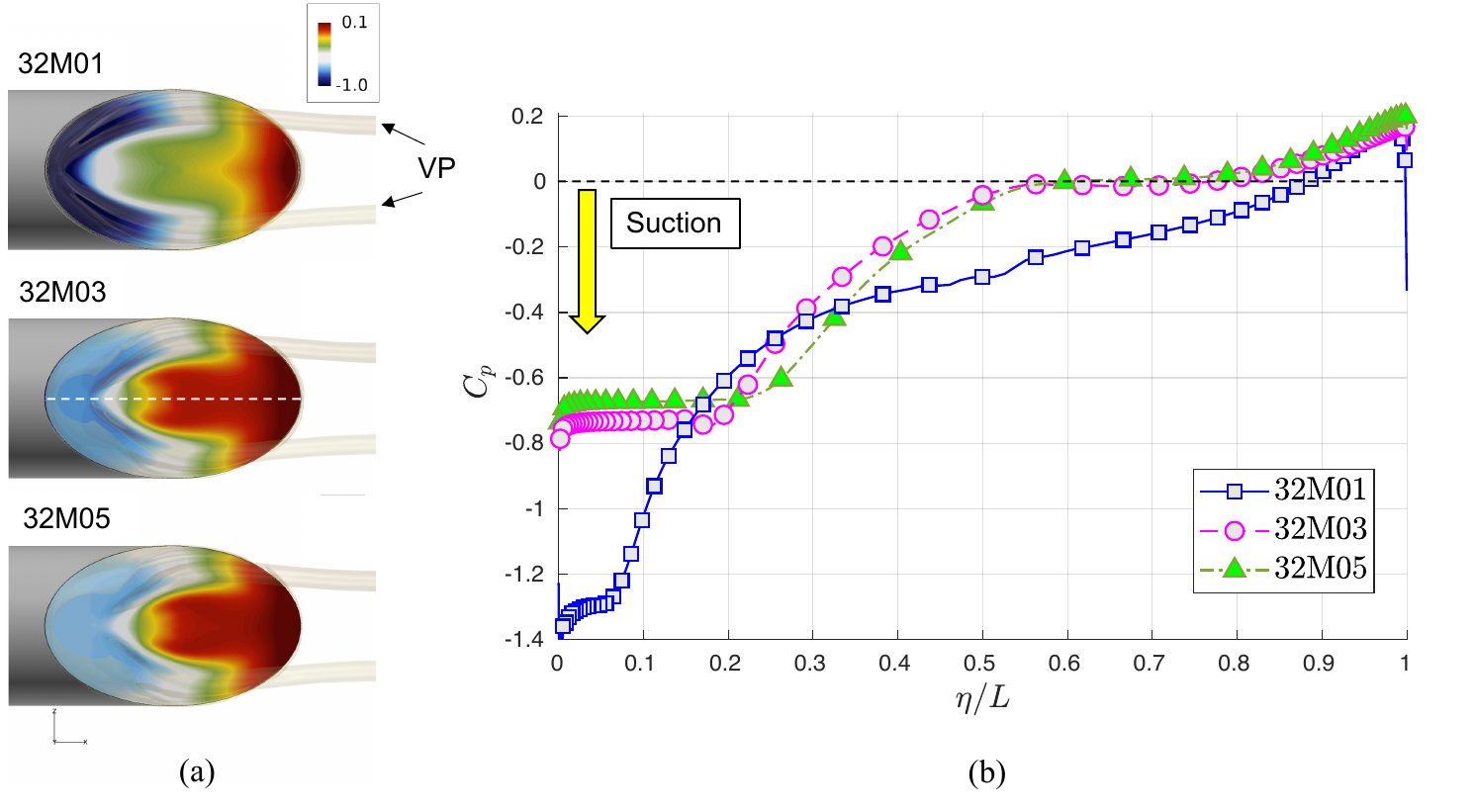}
    \caption{Sequence A: Effect of Mach number on the $C_p$ at the base (left) and $C_p$ distribution along the symmetry plane (right).}
    \label{fig:32Cp}
\end{figure}

%Another practical manifestation of the change in vortex evolution with compressibility is the 
%The change in the structure of the vortex pair is closely connected to the pressure footprint on the base. 
The changes in the recirculation region and vortex-pair development have a direct impact on the pressure distribution over the upswept base.
Figure~\ref{fig:32Cp}a shows the mean pressure coefficient, $C_p=(p-p_\infty)/(0.5\rho_\infty U_\infty^2)$, for the three Mach numbers.
The contours are plotted using identical levels to permit a direct comparison, and the corresponding vortex structures are superimposed in light gray for reference.
%{
%Figure~\ref{fig:32Cp}a shows the non-dimensional mean pressure distribution in terms of $C_p$ over the flat base for the three cases.
%This is reminiscent of vortex shedding in a bluff-body w
%Two major differences. Wake behind the geometry. The black line shows u=0. Large recirculation zone behind rounded geometry. Maybe connect to hysteresis/ wake flow-field.
%The angle of the downstream turbulent structures. Higher for the baseline.
%Here $C_p$ is the coefficient of pressure calculated as $C_p= (p-p_\infty)/(0.5 \rho_\infty U_\infty^2)$.
%The non-dimensional pressure distribution on the surface for the three Mach numbers is displayed in Fig.~\ref{fig:32Cp}.
%}
%{
%The contours are plotted using identical levels to permit a direct
%comparison, and the corresponding mean vortex structures are
%superimposed for reference.

%The vortex pairs are also highlighted for reference.
%At all Mach numbers, a horseshoe shaped suction footprint is observed on
%The ho
%A more quantiative variation is provided in 
%Figure~\ref{fig:32Cp}b shows the $C_p$ variation with $\eta/L$ over the $z=0$ symmetry plane.
%}
At Mach~0.1, the horseshoe vortex produces a pronounced U-shaped suction footprint ($C_p < 0$) over the base.
Similar footprints have been observed in previous studies of slanted-base afterbodies~\citep{bulathsinghala2017afterbody,zigunov2020reynolds,ranjan2020mean} and are associated with the head and developing legs of the horseshoe vortex.
The pressure deficit is particularly strong near the upstream apex, where the separated flow turns back toward the base and forms the
recirculation region.

Increasing the Mach number substantially weakens and redistributes this suction footprint.
The downstream movement of the vortex head is accompanied by a corresponding downstream displacement of the low-pressure region.
The most pronounced change occurs between Mach~0.1 and Mach~0.3; the pressure distributions for 32M03 and 32M05 are comparatively similar.
%{
%The changes in vortex formation and evolution discussed above have a
%direct aerodynamic manifestation through the pressure distribution on
%the upswept base.
%This suction region related to drag~\cite{}.
%This suction region spans the approximate area where 
%This U-shaped distribtuion has been studied extensively 
%}
A more quantitative comparison along the $z=0$ centerline is provided in Fig.~\ref{fig:32Cp}b.
The nearly constant-$C_p$ region immediately downstream of the upstream apex broadens considerably between 32M01 and the two higher-Mach-number cases.
This region approximately corresponds to the centerline trace of the recirculation zone and is consistent with its elongation observed in Fig.~\ref{fig:32SepBubble}.
The longer recirculation region allows the separated flow to turn more gradually toward the base, thereby reducing the local suction peak.
The minimum centerline value changes from approximately $C_p=-1.4$ for 32M01 to $C_p=-0.65$ for 32M05.
The 32M03 and 32M05 centerline distributions remain similar over most of the base and both recover to the freestream pressure, $C_p=0$, near $\eta/L\sim0.6$.
In contrast, the stronger pressure deficit in 32M01 persists to approximately $\eta/L\sim0.87$.
The weaker and less extensive pressure deficit in the 32M03 and 32M05 cases therefore indicate a smaller base-pressure-drag coefficient than for 32M01.
The similarity between the 32M03 and 32M05 distributions further indicates that most of the compressibility-induced change in the base-pressure loading occurs between Mach~0.1 and Mach~0.3.

%\subsection{Vortex-pair Evolution with Mach number}

%\begin{figure}
%    \centering
%    \subfloat[]{\includegraphics[width=0.9\textwidth]{Compressibility_Effects/Figures/MeanQ_streamlines_M01.pdf}} \\
%    \subfloat[]{\includegraphics[width=0.9\textwidth]{Compressibility_Effects/Figures/MeanQ_streamlines_M03.pdf}} \\
%     \subfloat{\includegraphics[width=0.9\textwidth]{Compressibility_Effects/Figures/MeanQ_streamlines_M05.pdf}}
%    \caption{Caption}
%    \label{fig:my_label}
%\end{figure}

%\begin{figure}
%    \centering
%    \includegraphics[width=0.7\textwidth]{Compressibility_Effects/Figures/Vortevolution.pdf}
%    \caption{Caption}
%    \label{fig:my_label}
%\end{figure}

%\subsection{Evolution of base surface features}

%\begin{figure}
%    \centering
%    \subfloat[]{\includegraphics[width=0.33\textwidth,trim=300 0 250 0, clip]{Compressibility_Effects/Figures/CUTCYL32_M01_basestreamlines_wCp.jpeg}}
 %    \subfloat[]{\includegraphics[width=0.33\textwidth,trim=300 0 250 0, clip]{Compressibility_Effects/Figures/CUTCYL32_M03_basestreamlines_wCp.jpeg}}
  %    \subfloat[]{\includegraphics[width=0.33\textwidth,trim=300 0 250 0, clip]{Compressibility_Effects/Figures/CUTCYL32_M05_basestreamlines_wCp.jpeg}}
   % \caption{Caption}
   % \label{fig:my_label}
%\end{figure}

%\clearpage
%\section{\label{sec:IV} Sequence B: Compressibility effects over ${45^\circ}$ aftbody}
\section{\label{sec:IV} Sequences B \& C: Flow over ${\textbf{45}^\circ}$ afterbody} \label{sec:45}

\subsection{Vortex-Regime and transition to Wake-closure}
\begin{figure}
    \centering
    %\subfloat[]{    
    \includegraphics[width=\textwidth]{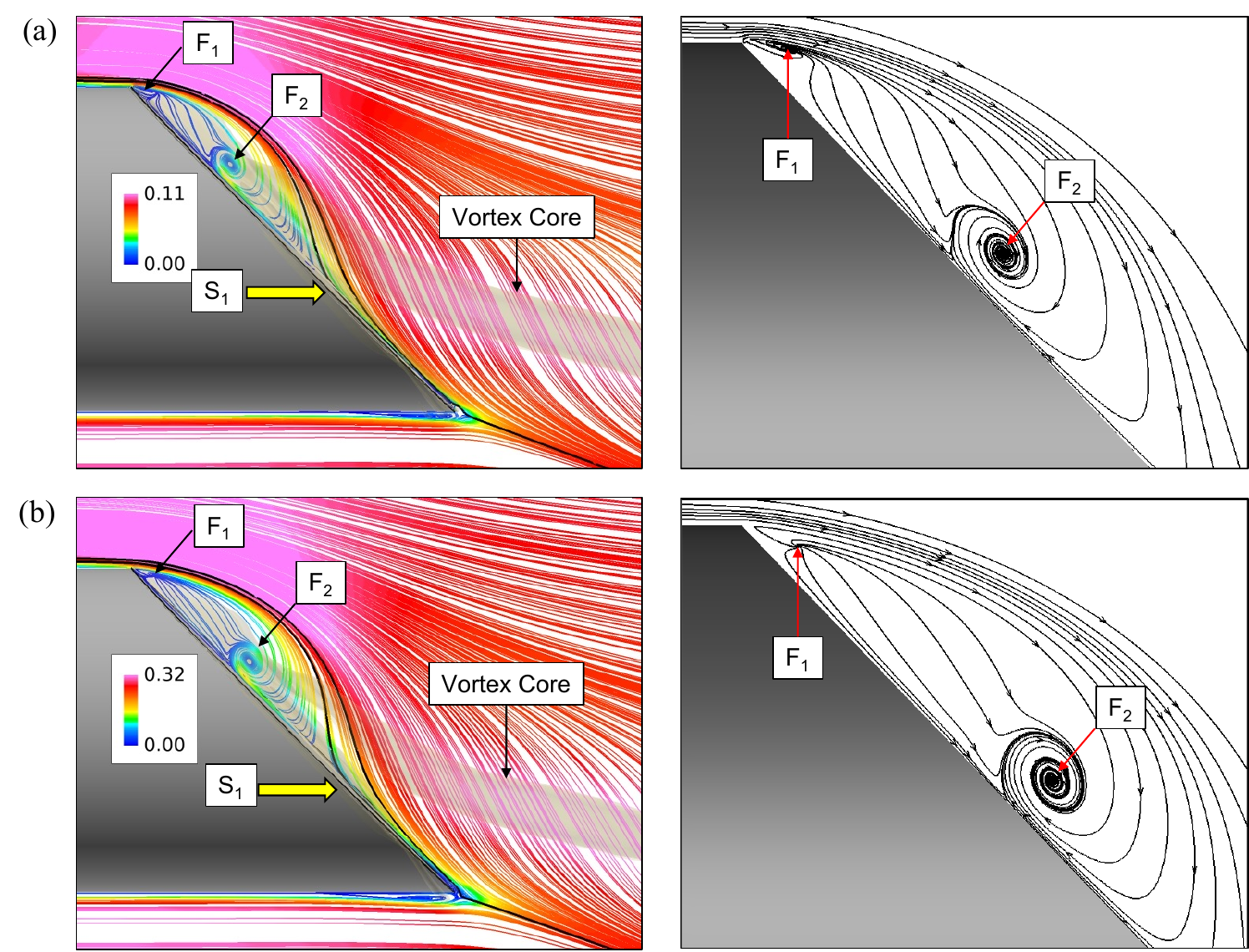}%}\\%\hspace{0.02\textwidth}
    %\subfloat[]{\includegraphics[width=\textwidth]{Compressibility_Effects/Figures/45M03_2DStreamlines_v2.pdf}} \\
    
   % \subfloat[]{\includegraphics[width=0.76\textwidth]{Compressibility_Effects/Figures/45M05_2DStreamlines.pdf}} \\
   \caption{Sequence B1: Streamlines in the symmetry plane for (a) 45M01V and (b) 45M03V. The vortex core is highlighted using $Q=37$ iso-surface.}
    \label{fig:45streamlines}
\end{figure}

We now consider the $45^\circ$ afterbody. %, for which the vortex-pair and separated-wake states are both known to be admissible at the baseline Mach number. 
Figure~\ref{fig:45streamlines} shows surface-restricted streamlines colored with Mach number on the symmetry plane for the cases 45M01V and 45M03V.
The streamlines are colored by the local Mach number, and the three-dimensional vortex core is superimposed using a $Q=37$ isosurface.
The right panels provide enlarged views of the symmetry-plane topology near the upstream apex.

%For reference, the vortex core is also highlighted using a $Q=37$ isosurface. 
%An enlarged view of the separation bubble is provided in the right insets of Figs~\ref{fig:45streamlines}a and~\ref{fig:45streamlines}b.
%As noted previously in $\S$~\ref{sec:LES}, compared to the $32^\circ$ afterbody, the $45^\circ$ configuration has larger separation bubble at Mach~0.1.
The flow topology shows some similarities with the 32M01 and 32M03 counterparts.
For instance, the two focii F$_1$ (source) and F$_2$ (sink) are clearly visible inside the separation bubble. 
The flow enters the symmetry plane through F$_1$ and leaves the symmetry plane via F$_2$. %being the sink.
The trace of the vortex head on the symmetry plane coincides with the node F$_2$. %inside the separation bubble.
As expected, increasing the Mach number from 0.1 to 0.3 produces a clear downstream extension of the recirculation region.
The position of F$_1$ remains close to the geometrically fixed
separation edge, whereas F$_2$ and the vortex head move farther
downstream.
The two black streamlines in Fig.~\ref{fig:45streamlines} further illustrate the increase in the size of the recirculation region with Mach number.
The lower streamline is the limiting feature that delineates the outer boundary of the recirculating flow.
Its reattachment point S$_1$ moves downstream from $\eta/L=0.56$ for 45M01V to $\eta/L=0.67$ for 45M03V, demonstrating the increase in recirculation-bubble length at Mach~0.3, similar to results at $\phi=32^o$.
The second streamline originates from the same location near the upstream apex in both cases.
For 45M01V, this streamline reattaches well prior to the downstream apex, whereas for 45M03V it remains separated over nearly the entire base and only grazes the surface near the downstream apex.
Thus, as observed for sequence~A, the principal effect of increasing Mach number is to delay reattachment and the subsequent formation of the horseshoe vortex.
%\begin{figure}
%    \centering
%    \subfloat[]{\includegraphics[width=\textwidth]{Compressibility_Effects/Figures/45M01V_streamlines3D_vortex.pdf}}\\
 %   \subfloat[]{\includegraphics[width=\textwidth]{Compressibility_Effects/Figures/45M03V_streamlines3D_vortex.pdf}}\\
  % %   \subfloat[]{\includegraphics[width=\textwidth]{Compressibility_Effects/Figures/45M05_streamlines3D_vortex.pdf}}\\
  %  \caption{Sequence B$_{2}$: Flow topology }
   % \label{fig:enter-label}
%\end{figure}

\begin{figure}
    \centering

   \subfloat[]{\includegraphics[width=0.49\textwidth,trim=80 220 100 220,clip]{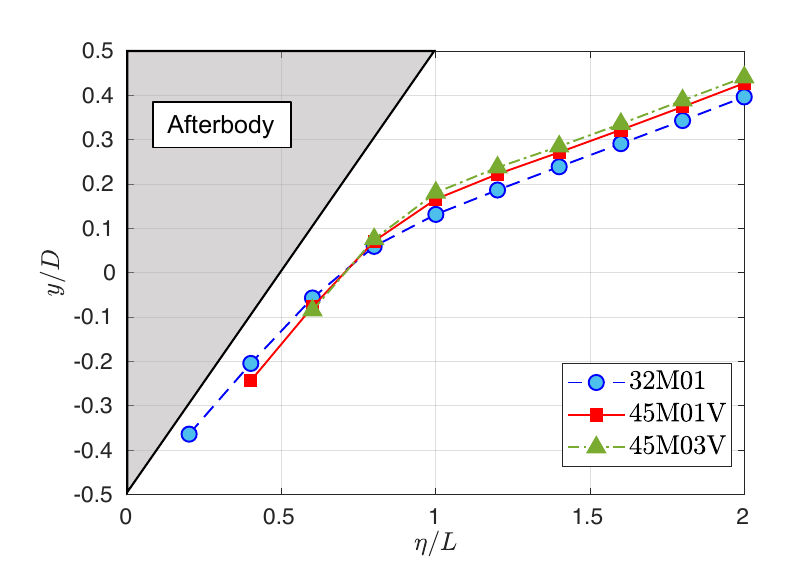}} \hspace{0.02in} 
   \subfloat[]{\includegraphics[width=0.49\textwidth,trim=80 220 100 220,clip]{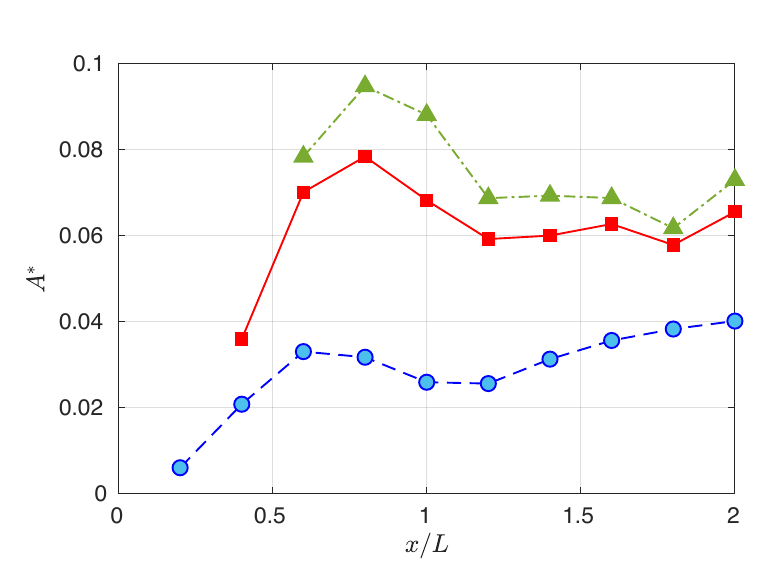}}
    \caption{Comparison of (a) vertical vortex location and (b) cross-sectional area for cases 45M01V and 45M03V.}
    \label{fig:45Grafiteaux}
\end{figure}

The downstream displacement of the vortex head is quantified in Fig.~\ref{fig:45Grafiteaux} using the $\Gamma_1/\Gamma_2$ approach.
The vertical vortex-core location and cross-sectional area are evaluated at intervals of $\Delta\eta=0.2L$.
Results for 32M01 are included to distinguish the effects of upsweep angle from those of Mach number.
At Mach~0.1, the first identifiable vortex core occurs farther downstream for the $45^\circ$ afterbody than for the $32^\circ$ afterbody; this result is consistent previous findings of~\cite{ranjan2020mean} and~\cite{zigunov2020reynolds}.
The additional increase in Mach number delays the vortex inception further, with the first identifiable 45M03V core occurring downstream
of that for 45M01V.
Once formed, however, the vortex cores in 45M01V and 45M03V follow similar vertical trajectories.
The close agreement between their $y/D$ locations indicates that compressibility primarily changes where the vortex pair begins to form, rather than substantially altering its subsequent mean trajectory in the vortex-pair regime.
The cross-sectional areas show a stronger Mach-number dependence.
The vortices over the $45^\circ$ afterbody are larger than those in 32M01, and the 45M03V vortices are consistently larger than their 45M01V counterparts after inception.
Thus, while increasing Mach number delays the formation of the vortex pair, the resulting vortices again undergo rapid downstream growth.
%{
%In order to quantify the effect of compressibility on the vortex pair, Figs.~\ref{fig:45Grafiteaux}a and~\ref{fig:45Grafiteaux}b show the vertical vortex location and the vortex cross-sectional area at every $\Delta \eta = 0.2L$ using the $\Gamma_1/\Gamma_2$ approach.
%The results for the 32M01 case are also re-plotted here for reference.
%Consistent with previous findings~\cite{ranjan2020mean,zigunov2020reynolds}, the vortex core is formed 
%Compared to the 32M01 case, the vortex core is formed much downstream for the 45M01V case, consistent with previous findings of ~\cite{ranjan2020mean} and ~\cite{zigunov2020reynolds}. 
%As seen previously in Fig.~\ref{}, with an increase in Mach number
%Despite these differences, when scaled with $D$ and $L$, the vertical locations show a reasonable collapse for both $32^\circ$
%Moreover, the vortex cores in the 45M01V case are formed much closer to the flat base, resulting in a stronger induced drag as shown later. 
%As expected, with an increase in Mach number, the formation of the horseshoe vortex is pushed further downstream. 
%Despite this difference, the vortex vertical locations show a reasonable collapse for 45M01V and 45M03V cases.
%With an increase in Mach number, the vortex core is pushed even closer the  
%Area larger than 32. %Further increase with Mach number.
%}

\begin{figure}
    \centering
    \includegraphics[width=\textwidth]{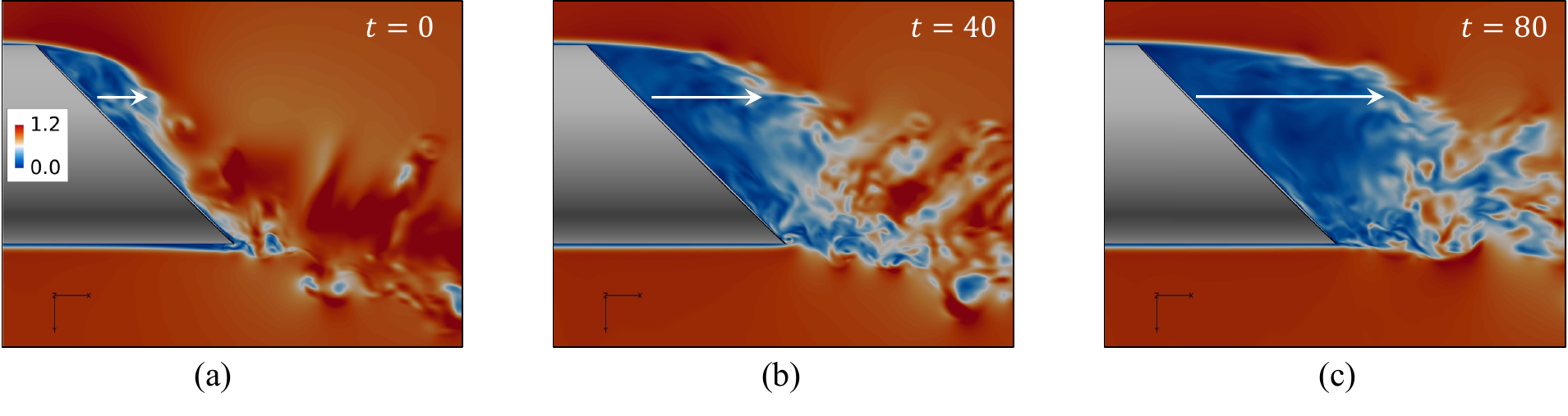}
    \caption{Sequence B2: Velocity magnitude contours showing transition of flow from vortex to wake regime with time.}
    \label{fig:vorttowake}
\end{figure}

The gradual downstream extension of the recirculation region between Mach~0.1 and Mach~0.3 culminates in a qualitative change in flow state when the Mach number is increased to 0.5.
Figure~\ref{fig:vorttowake} shows the temporal development of this transition using instantaneous velocity-magnitude contours.
At $t=0$, immediately following the increase to Mach~0.5, the flow retains the inherited vortex-pair structure of 45M03V.
By $t=40$, the low-velocity recirculating region has expanded across essentially the entire upswept surface, and the organized vortex-head structure has substantially weakened.
At $t=80$, the transition is complete: the flow no longer closes
through a coherent streamwise vortex pair and instead exhibits a
large separated-wake region extending beyond the downstream apex.

At the same Mach number, the recirculation region is already larger for the $45^\circ$ afterbody than for the $32^\circ$ afterbody because of the greater upsweep angle.
The additional compressibility-induced growth at Mach~0.5 causes the separated region to extend across the entire upswept base.
This progression closely parallels the wake-state transition produced by increasing upsweep angle in previous studies.
\cite{ranjan2020mean} documented the evolution from increasingly large horseshoe-vortex structures to a wake-type flow at
$\phi=55^\circ$, while~\cite{zigunov2020reynolds} conjectured that the transition occurs when the separation-bubble length becomes comparable to the length of the slanted surface.
In those investigations, the growth of the separated region was
produced by a change in geometry or Reynolds number.
Here, the upsweep angle and $Re_D$ remain fixed; compressibility
provides the route by which the same limiting state is reached.
A similar transition was observed in the recent experiments of~\cite{huss2025compressibility}, who observed a corresponding change from a vortex-dominated state at Mach~0.3 to a fully separated wake at Mach~0.6 for a sharp-edged $45^\circ$ afterbody.

\begin{figure}
    \centering
    %\subfloat[]{\includegraphics[width=\textwidth]{Compressibility_Effects/Figures/45M03V_streamlines3D_vortex.pdf}} \\
   %\subfloat[]{\includegraphics[width=\textwidth]{Compressibility_Effects/Figures/45M05_streamlines3D_vortex.pdf}}\\
   \includegraphics[width=\textwidth]{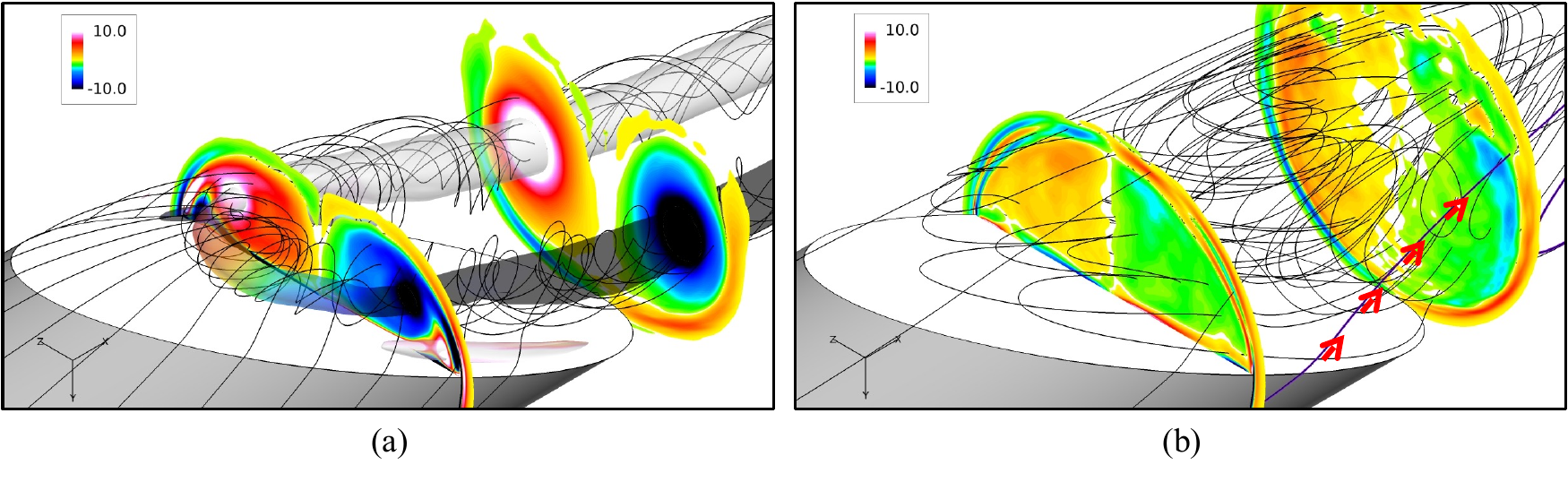}
    \caption{Vortex Pair vs Wake Regime: 45M05 Flow topology for 45M03V (left) and 45M05W (right).}
    \label{fig:45M05}
\end{figure}

In the vortex-pair state, the separation bubble closes on the upswept
base, and its recirculating flow turns smoothly into the head and legs
of the horseshoe vortex.
When the recirculation region extends to the downstream apex, the
reattachment and roll-up process can no longer be completed on the
base.
The organized streamwise vortex-pair formation pathway is consequently
replaced by a broad separated wake.
The distinction between the two states is further illustrated in
Fig.~\ref{fig:45M05}. 
In 45M03V, the streamlines wrap around two coherent, counter-rotating
vortex tubes (visualized by isolevels of $Q$-criterion), and the crossflow planes contain concentrated regions of
opposite-signed streamwise vorticity.
In 45M05W, these compact vortex cores are absent -- contours of $Q$-criteria are incoherent and thus not shown.
The streamwise vorticity is instead distributed over broader,
less-organized structures, while the streamlines pass through the
large recirculating wake rather than winding around a persistent
vortex pair.

%{
%The corresponding symmetry-plane streamlines are shown in Fig.~\ref{fig:45M05streamlines}.
%\begin{figure}
%   \centering
%    \includegraphics[width=0.76\textwidth]{Compressibility_Effects/Figures/45M05_2DStreamlines.pdf} 
 %   \caption{45M05: Streamlines in the symmetry plane.}
  %  \label{fig:45M05streamlines}
   % \end{figure}
%Unlike the compact recirculation region of the vortex-pair state, the 45M05W separated region occupies the full streamwise extent of the upswept surface and extends into the near wake.
%The mean topology therefore resembles the fully separated wake
%reported previously when the critical upsweep angle was exceeded \cite{morel1978effect,morel1980effect,britcher1991interference, ranjan2020mean}.
%The important distinction is that the present transition is produced without changing either the geometry or Reynolds number: increasing Mach number alone causes the $45^\circ$ flow to leave the vortex-pair branch and adopt the separated-wake state.
%}

%\clearpage
%\clearpage
\subsection{\label{sec:Hysteresis} Hysteresis and Drag Implications}
%\subsection{Persistence of separated-wake state}
Hysteresis between the vortex-pair and separated-wake states has previously been demonstrated by varying either the upsweep angle or Reynolds number.
At fixed Reynolds number, increasing and subsequently decreasing
$\phi$ can produce different flow states at the critical angle near
$45^\circ$~\citep{morel1980effect,ranjan2020hysteresis}.
Similarly, Reynolds-number sweeps at fixed $\phi=45^\circ$ have shown
that the transition from the separated-wake state to the vortex-pair
state occurs at a different $Re_D$ than the reverse transition~\citep{ranjan2020hysteresis,
zigunov2022hysteretic}.
Thus, over a finite range of conditions, either state may be sustained
depending on the path by which the final condition is reached.

\begin{figure}
    \centering
    \includegraphics[width=0.8\textwidth]{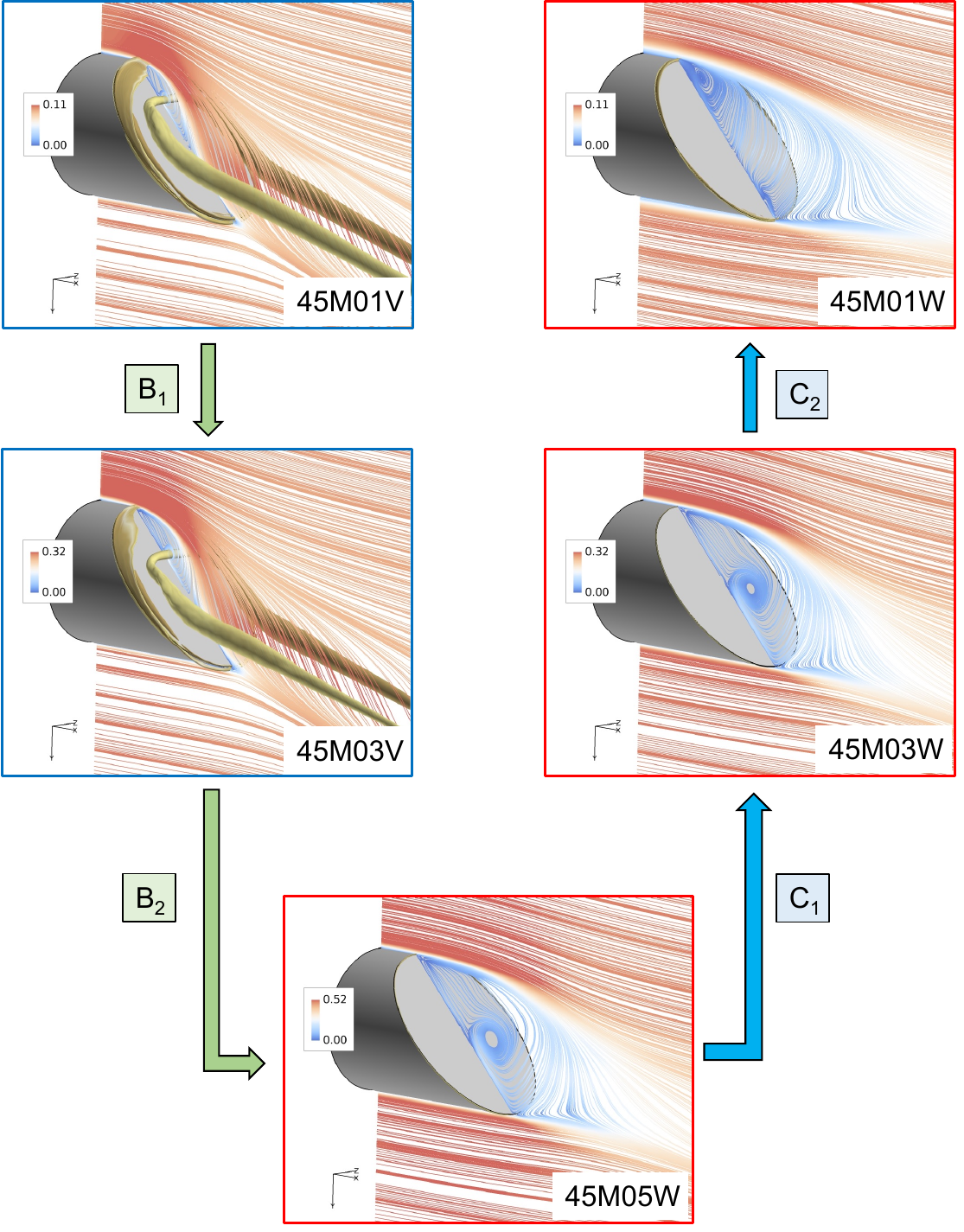}
    \caption{$45^\circ$ Afterbody: Schematic showing the switch from a vortex-pair state to a separate-wake state and the persistence of the separated-wake state at low Mach numbers. }
    \label{fig:Hysteresis}
\end{figure}

The present simulations demonstrate an analogous history dependence
with Mach number.
As summarized in Fig.~\ref{fig:Hysteresis}, sequence~B begins from the 45M01V vortex-pair state.
Increasing the Mach number ($B_1$) first produces the 45M03V state and then causes a transition ($B_2$) to the separated-wake state at Mach~0.5.
Sequence~C is initiated from this 45M05W state and reverses the Mach-number progression.
When the Mach number is reduced to 0.3 ($C_1$), the flow remains in the separated-wake state, yielding 45M03W rather than returning to 45M03V.
A further reduction to Mach~0.1 ($C_2$) similarly produces 45M01W, even though the vortex-pair state 45M01V is also supported at the same Mach number.

%{
%Starting from 45M05 wake regime, as the Mach number is reduced to 0.3, wake regime persists.
%The streamlines resemble the 45M05 case, the separation bubble still envelopes the upswept base. 
%On further reducing to 0.1, we get the wake regime obtained by Ranjan and Zigunov by reducing Re.
%The effect of Re is discussed in $\S$~\ref{sec:Re_effects}
%No switching back from vortex to wake, simulations run for long time.
%\begin{figure}
%    \centering
%    \subfloat[]{\includegraphics[width=0.48 \textwidth]{Compressibility_Effects/Figures/45M01W_2DStreamlines.pdf}} \hspace{0.02\textwidth}
 %   \subfloat[]{\includegraphics[width=0.48 \textwidth]{Compressibility_Effects/Figures/45M03W_2DStreamlines.pdf}}
  %  \caption{Separated-wake state: 2D Streamlines on the symmetry plane for 45M01W and 45M03W cases.}
   % \label{fig:my_label}
%\end{figure}
%}

The distinction between the two branches is evident from their mean
flow topology.
Throughout sequence~C, the recirculation region continues to occupy the full streamwise extent of the upswept base, and no coherent streamwise vortex pair re-forms.
The symmetry-plane streamline patterns for 45M03W and 45M01W therefore remain qualitatively similar to that of 45M05W rather than to their vortex-branch counterparts.
Each reverse-sequence simulation was advanced for more than $300D/U_\infty$ after the initial transients, exceeding those of previous hysteresis studies~\citep{ranjan2020hysteresis}.
No spontaneous transition from the separated-wake state back to the
vortex-pair state was observed.
The persistence of the two states at identical values of $\phi$, $Re_D$, and Mach number demonstrates hysteresis with Mach number, rather than a temporary delay in the flow response.

\begin{figure}
    \centering
    \includegraphics[width= \textwidth]{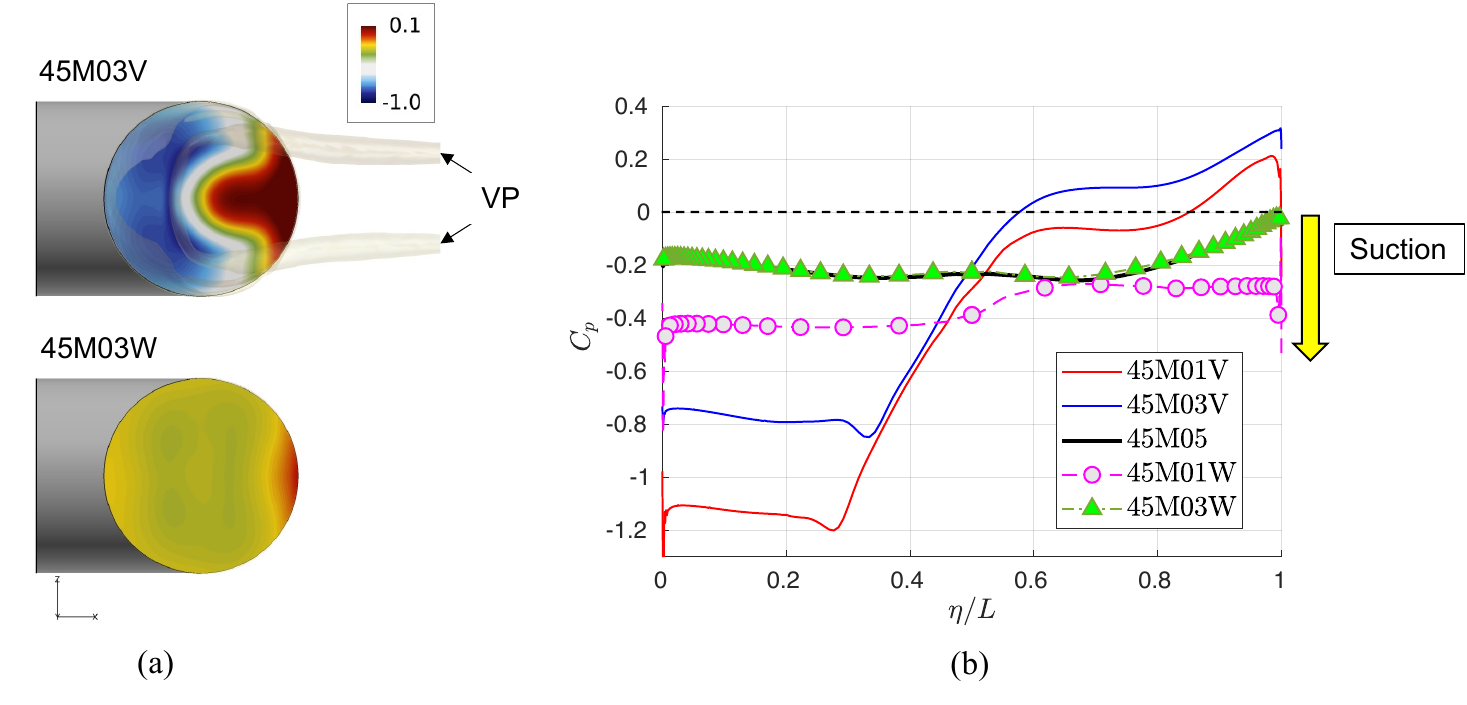}
    
    \caption{Effect of hysteresis on the $C_p$ variation over the base.}
    \label{fig:45Cp}
\end{figure}

%\subsection{Base pressure}
The transition between the vortex-pair and separated-wake states also produces a substantial change in the pressure loading over the upswept base.
Figure~\ref{fig:45Cp} compares the mean $C_p$ distributions for all
cases in sequences B and C.
The corresponding centerline profiles provide a quantitative
comparison of their streamwise variation.
In the vortex-pair states 45M01V and 45M03V, the pressure field
contains the pronounced horseshoe-shaped suction footprint associated
with the vortex head and its developing legs.
The pressure varies strongly over the base, with localized regions of
low $C_p$ beneath the vortical structures.
Increasing the Mach number from 0.1 to 0.3 displaces this footprint
downstream, consistent with the enlargement of the recirculation
region and the downstream movement of the vortex head discussed
previously.

The transition to 45M05W produces a qualitatively different pressure
field.
The localized horseshoe-shaped suction footprint is no longer present.
Instead, the separated-wake state exhibits a broader and more nearly
uniform pressure deficit over the base.
The corresponding centerline distribution remains comparatively flat
over much of the upswept surface before changing near the downstream
apex.
This behavior is characteristic of the separated-wake state and is related to the large, slowly turning recirculation
region that occupies the region beneath the base~\citep{ranjan2020hysteresis}.

The pressure distribution retains this wake-like character as the Mach number is reduced in sequence~C.
Both 45M03W and 45M01W exhibit the broad, relatively uniform pressure field of 45M05W rather than recovering the localized suction footprint of the corresponding vortex-pair states.
Consequently, at Mach~0.3, the 45M03V and 45M03W cases have distinct surface-pressure distributions despite having identical values of Mach number, Reynolds number, and upsweep angle.
The same distinction is observed between 45M01V and 45M01W at Mach~0.1. 
As summarized in Table~\ref{tab:45_drag}, the separated-wake state has a lower base-pressure drag coefficient than the corresponding vortex-pair state by 0.199 at Mach~0.3 and by 0.161 at Mach~0.1.
\begin{table}
    \centering
    \caption{Base-pressure drag coefficients for the vortex-pair (VP) and separated-wake states of the $45^\circ$ afterbody.}
    \label{tab:45_drag}
    \begin{tabular}{cccc}
        \hline
        $M_\infty$
        & VP
        & Wake
        & $\Delta C_{D,b}$\\
        \hline
        0.1 & 0.502 & 0.341 & 0.161 \\
        0.3 & 0.349 & 0.150 & 0.199 \\
        0.5 & ---   & 0.156 & ---   \\
        \hline
    \end{tabular}
\end{table}
Along the separated-wake branch, most of the Mach-number dependence occurs between Mach~0.1 and Mach~0.3, over which $C_{D,b}$ decreases from 0.341 to 0.150.
The centerline pressure distributions of 45M03W and
45M05W are nearly coincident, and their base-pressure drag coefficients are correspondingly similar.

The base-pressure response is therefore multivalued with Mach number and mirrors the hysteresis observed in the mean flow topology.
The loading on the upswept surface depends not only on the instantaneous Mach number but also on whether that condition is approached from the vortex-pair or separated-wake branch.

\subsection{Sensitivity to $Re_D$} \label{sec:Re_effects}
Reynolds-number effects require particular consideration for the $45^\circ$ afterbody.
\cite{ranjan2020hysteresis} showed that, under incompressible conditions, reducing $Re_D$ can independently drive the flow from the vortex-pair state to the separated-wake state.
To determine whether the wake retained along sequence~C is confined to the baseline Reynolds number, the statistically stationary 45M03W solution at $Re_D=25{,}000$ was used to initialize an additional calculation at $Re_D=75{,}000$, while retaining
$M_\infty=0.3$ and $\phi=45^\circ$.
The higher-Reynolds-number calculation employed sixth-order spatial discretization.
Despite the threefold increase in $Re_D$, the flow remains in the separated-wake state and does not recover an organized vortex pair.

\begin{figure}
    \centering
    \includegraphics[width=0.6\linewidth,trim=100 280 120 280,clip]{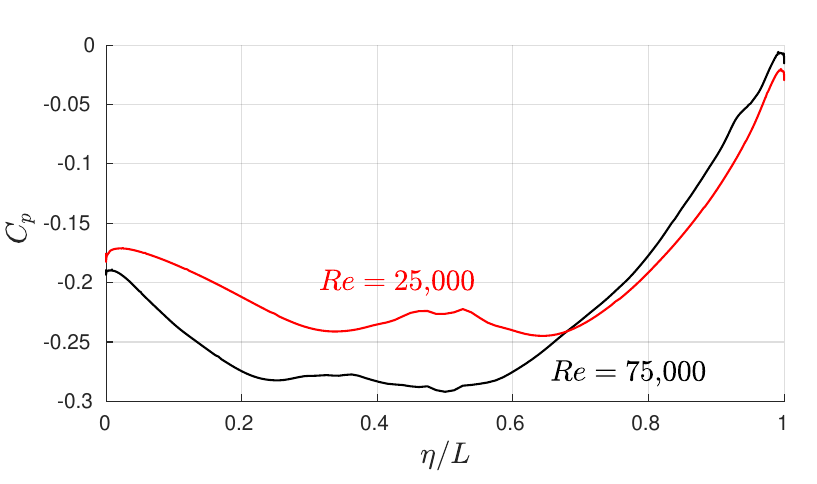}
    \caption{Comparison of centerline $C_p$ distributions for the
45M03W separated-wake state at $Re_D=25{,}000$ and $75{,}000$.}
    \label{fig:45ReCp}
\end{figure}

Figure~\ref{fig:45ReCp} compares the corresponding centerline pressure
distributions.
Both cases retain the broad pressure deficit characteristic of the separated-wake state, demonstrating that the wake selected through the Mach-number sequence persists at the higher Reynolds number.
However, the pressure levels are not identical. 
The $Re_D=75{,}000$ case exhibits lower $C_p$ values over much of the upstream and central portions of the base, while the two distributions approach one another toward the downstream apex.
Reynolds number therefore modifies the quantitative pressure loading
within the wake state without changing its mean-flow topology.

%\clearpage

%\section{\label{sec:Re_effects}Effect of Reynolds number}
%As stated previously in $\S$~\ref{sec:LES}, the Mach number was increased keeping the $Re_D$ constant.
%This assumed that the flow 
%The previous sections demonstrate that Mach number can be strategically used to lower drag.
%Previous investigation has shown that $Re$ can be used to switch from vortex to wake at Mach~0.1.
%Thus we have to establish $Re$ independence at higher Mach numbers. We demonstrate this for 45M03 case. 
%Since the freestream Mach number was increased to 0.3. 
%We increase the $Re$ to $7.5 \times 10^4$.
%Sixth-order compact scheme with eighth-order filter.
%Wake to vortex by lowering . 
%Increase speed of aircraft without increasing Mach number.

%\clearpage
\section{\label{sec:Conclusion} Conclusion}
The present investigation demonstrates that Mach number history governs the aerodynamic state of a highly upswept afterbody by controlling whether the separated shear layer rolls up into a streamwise vortex pair or remains as a broad separated wake.
To isolate the effect of compressibility, three sequences of wall-resolved large-eddy simulations were conducted at $M_\infty=0.1$, 0.3, and 0.5 for cylindrical aftbodies at upsweep angles of $32^\circ$ and $45^\circ$, while holding the Reynolds number fixed at $Re_D=25{,}000$.
Across both upsweep angles, the principal effect of increasing $M_\infty$ is the downstream growth of the recirculation region originating near the upstream apex. 
This growth delays reattachment and shifts the formation of the vortex head downstream, thereby changing how much of the upswept surface remains available for the separated shear layer to roll up into an organized streamwise vortex pair.

For the $32^\circ$ afterbody, the vortex-pair topology persists over
the full $M_\infty$ range.
Increasing $M_\infty$ shifts vortex-head formation downstream.
However, the vortex pair lifts off from the upswept surface at approximately the same axial location for all three Mach numbers. 
By this point, the higher-Mach-number vortices have attained larger core areas and greater circulation. 
Their spanwise separation remains nearly unchanged, whereas their vertical trajectories increasingly diverge downstream. Despite the stronger downstream vortices, the associated suction footprint on the base becomes substantially weaker because vortex development occurs farther downstream and away from the surface.

For the $45^\circ$ afterbody, the same growth of the recirculation region produces a regime transition.
At Mach~0.5, the separated region extends across the upswept base, preventing the reattachment and roll-up required to form a coherent streamwise vortex pair. 
The flow instead establishes a broad separated wake. Because the upsweep angle and $Re_D$ remain fixed throughout the sequence, compressibility provides the route to this transition. 
Moreover, the separated-wake state persists as $M_\infty$ is subsequently reduced to 0.3 and 0.1. 
The vortex-pair and separated-wake states can therefore occur at identical values of $M_\infty$, upsweep angle, and $Re_D$, depending on the Mach-number history. 
Relative to the vortex-pair state, the separated-wake state exhibits a broader, more uniform base-pressure distribution and correspondingly lower base-pressure drag.
An additional simulation at Mach~0.3 and $Re_D=75{,}000$ also retains the separated-wake topology, showing that the observed wake branch is not confined to the baseline $Re_D$.
%These results show that compressibility 

\section*{Acknowledgments}

This work was performed in part under the sponsorship of U.S. Air Force Office of Scientific Research with Dr. Gregg Abate serving as the project monitor. 
DVG also acknowledges partial support from the Collaborative Center for Aeronautical Sciences.
The views and conclusions contained herein are those of the authors and do not represent the opinion of AFOSR or the U.S. government. 
The authors are grateful for computational resource grants from the DoD HPCMP and the Ohio Supercomputer Center.
This work is declared Distribution A, approved for public release; distribution unlimited (PA Number AFRL-2026-3606).

\nocite{*}
%\clearpage
\bibliography{Ref}% Produces the bibliography via BibTeX.

\end{document}